\documentclass[journal]{IEEEtran}
\usepackage{bm}

\long\def\comment#1{}

\newfont{\bbb}{msbm10 scaled 700}

\newcommand{\ev}{{\bf e}}
\newcommand{\fv}{{\bf f}}

\newcommand{\sv}{{\bf s}}

\newcommand{\uv}{{\bf u}}

\newcommand{\xv}{{\bf x}}

\newcommand{\Am}{{\bf A}}

\newcommand{\Dm}{{\bf D}}

\newcommand{\Fm}{{\bf F}}

\newcommand{\Hm}{{\bf H}}
\newcommand{\Id}{{\bf I}}

\newcommand{\Lm}{{\bf L}}

\newcommand{\Um}{{\bf U}}

\newcommand{\Vm}{{\bf V}}

\newcommand{\Lam}{{\bf \Lambda}}

\newcommand{\Pim}{\hbox{\boldmath$\Pi$}}

\renewcommand{\arg}{{\hbox{arg}}}

\usepackage{paralist}
\usepackage{color}
\usepackage{graphicx}
\usepackage{tcolorbox}
\newenvironment{cframed}[1][blue]
  {\begin{tcolorbox}[colframe=#1,colback=white]}
  {\end{tcolorbox}}

\def\NEW#1{#1}%

\newtheorem{definition}{Definition}
\usepackage{tikz}
\usetikzlibrary{arrows.meta,positioning}

\ifCLASSINFOpdf
\else
\fi
\usepackage{ulem}

\usepackage{bm}
\usepackage{relsize}
\usepackage{amsmath,amssymb,amsfonts}
\begin{document}
%
\title{Graph Signal Processing: Overview, Challenges and Applications}
%
%
%

\author{Antonio~Ortega,~\IEEEmembership{Fellow,~IEEE,}
        Pascal~Frossard,~\IEEEmembership{Fellow,~IEEE,}
        Jelena~Kova\v{c}evi\'c,~\IEEEmembership{Fellow,~IEEE,}
        Jos\'e M.~F.~Moura,~\IEEEmembership{Fellow,~IEEE,}
       and~Pierre~Vandergheynst
\thanks{A. Ortega is with the University of Southern California, Los Angeles,
California, USA.}
\thanks{P.~Frossard and P.~Vandergheynst are with EPFL, Lausanne, Switzerland}
\thanks{J.~Kova\v{c}evi\'c and J.~M.~F.~Moura are with Carnegie Mellon University, Pittsburgh, Pennsylvania, USA}}

%
%

\markboth{Submitted to Proceedings of the IEEE}%
{Shell \MakeLowercase{\textit{et al.}}: Graph Signal Processing}
%



\maketitle

\begin{abstract}
Research in Graph Signal Processing (GSP) aims to develop tools for processing data defined on irregular graph domains. In this paper we first provide an overview of core ideas in GSP and their connection to conventional digital signal processing, \NEW{along with a brief historical perspective to highlight how concepts recently developed in GSP build on top of prior research in other areas}. We then summarize recent 
\NEW{advances} in developing basic GSP tools, including methods for sampling, filtering or graph learning. Next, we review progress in several application areas using GSP, including processing and analysis of sensor network data, biological data, and applications to image processing and machine learning. 
\end{abstract}




%
\IEEEpeerreviewmaketitle

\section{Introduction and motivation}
Data is all around us, and massive amounts of it. Almost every aspect of human life is now being recorded at all levels: from the marking and recording of processing inside the cells starting with the advent of fluorescent markers, to our personal data through health monitoring devices and apps, financial and banking data, our social networks, mobility and traffic patterns, 
marketing preferences, fads, and many more. The complexity of such networks \cite{newman2010networks} and interactions means that the data now resides on irregular and complex structures that do not lend themselves to standard tools.

Graphs offer the ability to model such data and complex interactions among them. For example, users on Twitter can be modeled as nodes while their friend connections can be modeled as edges. This paper explores adding attributes to such nodes and modeling those as signals on a graph; for example, year of graduation in a social network, temperature in a given city on a given day in a weather network, etc. Doing so requires us to extend classical signal processing concepts and tools such as Fourier transform, filtering and frequency response to data residing on graphs. It also leads us to tackle complex tasks such as sampling in a principled way. The field that gathers all these \NEW{questions} under a common umbrella is {\it graph signal processing (GSP)} \cite{SandryhailaM:13,ShumanNFOV:13}.

While the precise definition of a graph signal will be given later in the paper, \NEW{let us assume for now that} a graph signal is a set of values residing on a set of nodes. These nodes are connected via (possibly weighted) edges. As in classical signal processing, such signals can stem from a variety of domains; unlike in classical signal processing, however, the underlying graphs can tell a fair amount about those signals through their structure. Different types of graphs model different types of {\it networks} that these nodes represent. 

Typical graphs that are used to represent common real-world data include Erd\H{o}s-R\'enyi graphs, ring graphs, random geometric graphs, small-world graphs, power-law graphs, nearest-neighbor graphs, scale-free graphs, and many others. These model networks with random connections (Erd\H{o}s-R\'enyi graphs), networks of brain neurons (small-world graphs), social networks (scale-free graphs), \NEW{and many others.} 

As in classical signal processing, graph signals can have properties, such as smoothness, \NEW{that need to be} appropriately defined. They can also be represented via basic atoms and can have a spectral representation. In particular, the graph Fourier transform allows us to develop the intuition gathered in the classical setting and extend it to graphs; we can talk about the notions of frequency and bandlimitedness, for example. We can filter graph signals. They can be sampled, a notoriously hard problem; with graph signal processing, one gains access to principled tools mimicking the classical ones. We can denoise graph signals, we can learn their underlying structure, we can model them. If the graphs cannot be directly observed, we can also learn their structure from data. All of these topics will be explored in more detail in what follows.

As illustration, consider 
\NEW{what} smoothness of graph signals 
\NEW{may represent} in urban settings. First, however, we have to understand what smoothness means on graphs. For example, we can think of {\it smooth} graph signals in the vertex domain, that \NEW{are, signals where}
neighboring nodes tend to have similar values. We can also think of the smoothness of graph signals in the spectral domain, typically called bandlimitedness. Different types of smoothness are possible in the spectral domain where, instead of a sharp cut-off, frequency content may decay according to some law. 

\begin{figure}[htb]
  \begin{center}
    \begin{tabular}{cc}
 \includegraphics[width=0.45\columnwidth]{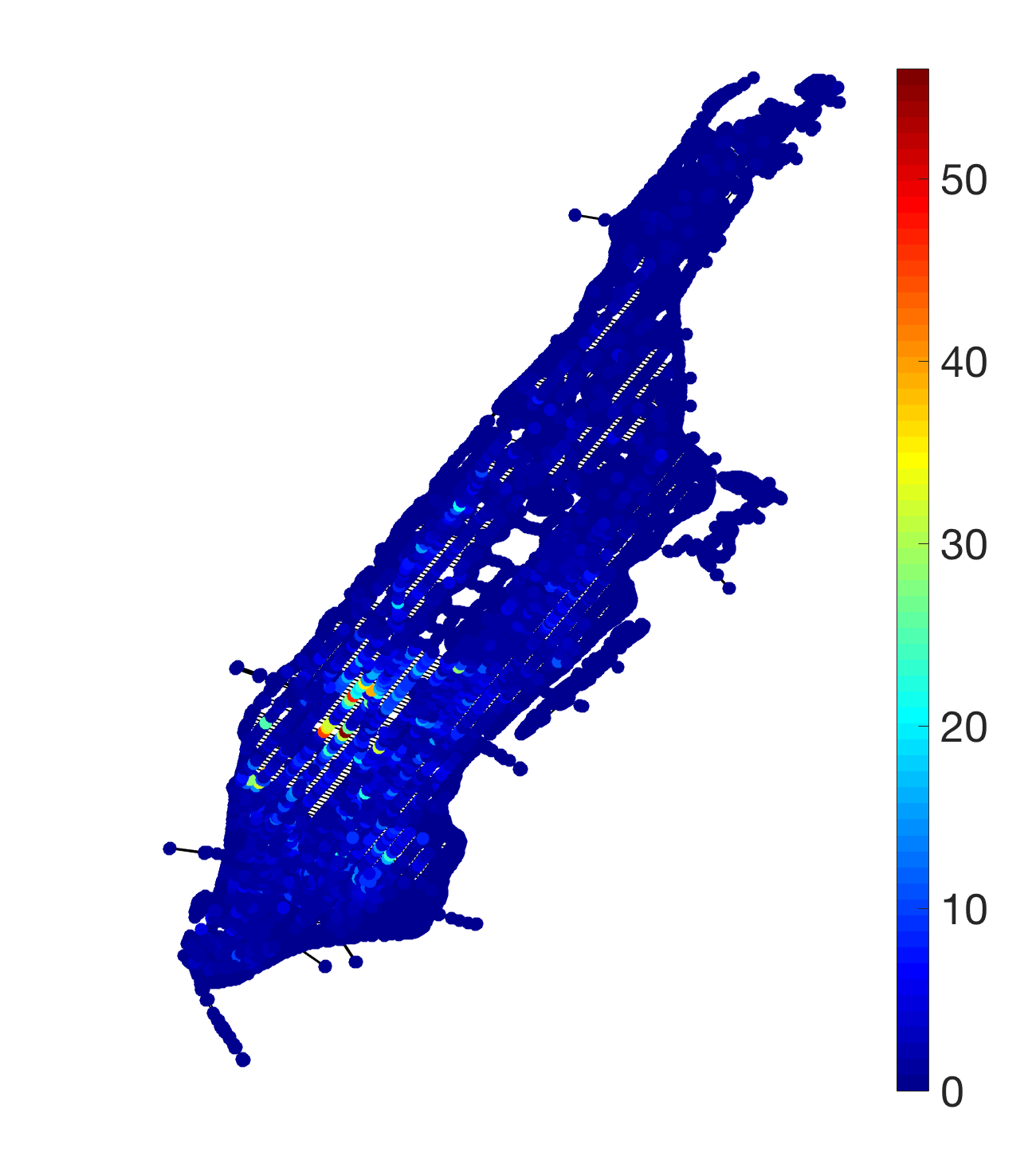} &
 \includegraphics[width=0.45\columnwidth]{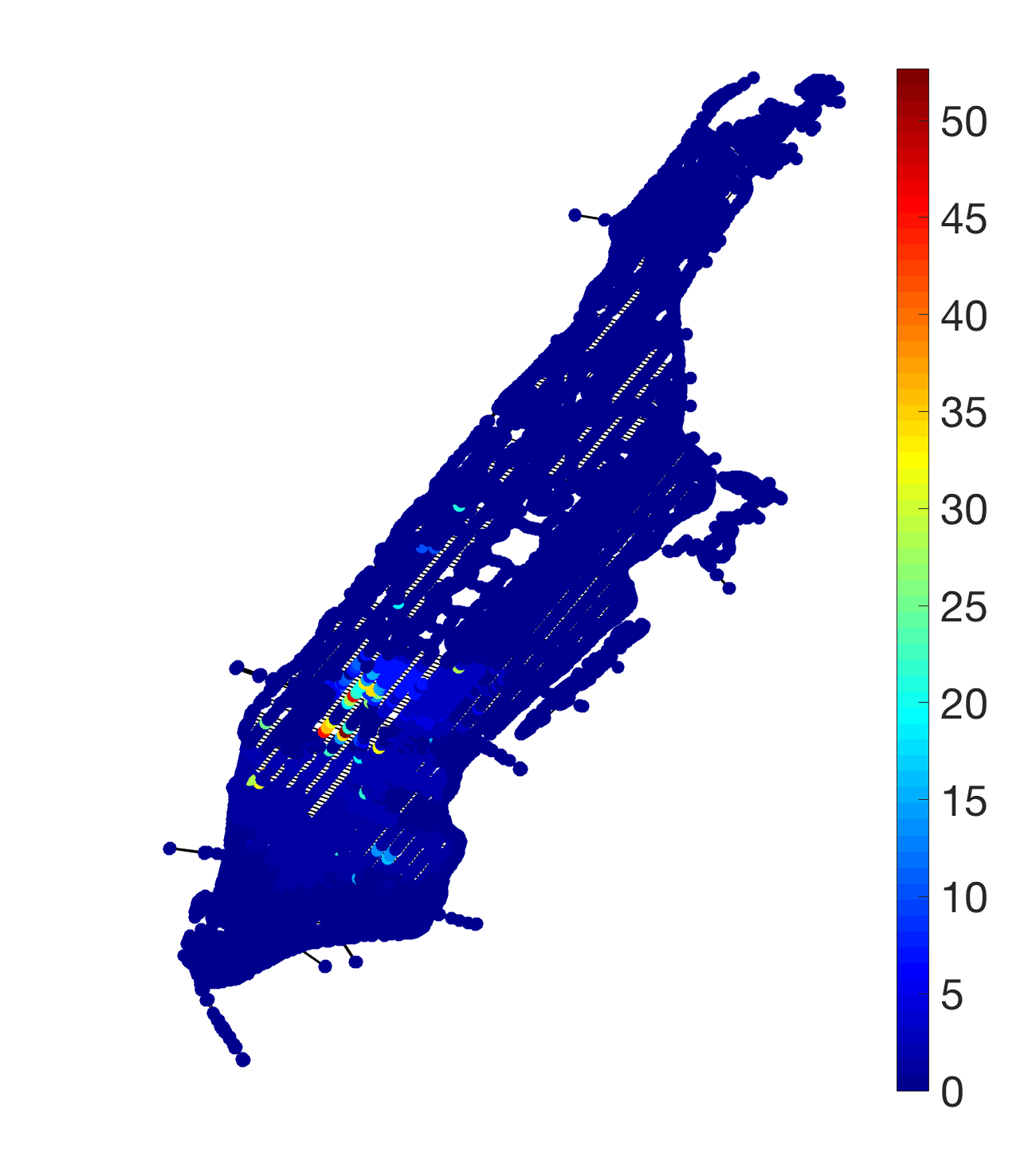} 
      \\
      {\small (a) Actual data } &  {\small (b) Approximation}
    \end{tabular}
  \end{center}
  \caption{\label{fig:toy_pc} Piecewise-smooth graph signals approximate irregular, non-smooth graph signals by capturing both large variations at boundaries as well as small variations within pieces. (a) Data captured in Manhattan (13,679 intersections), (b) Piecewise-smooth approximation to the data  with 50 coefficients (from~\cite{ChenSK:17} with permission).}
\end{figure}

Figure~\ref{fig:toy_pc} illustrates how a piecewise-smooth signal model can be used to approximate the taxi-pickup distribution in Manhattan. Figure~\ref{fig:toy_pc}(a) shows the number of taxi pickups (with blue as low numbers up to red for high numbers) from 6-7pm on January 4th, 2015, projected to the nearest intersection. What one can observe is that the busy shopping/entertainment areas such as Times Square in New York City show similar mobility patterns, as illustrated by the number of taxi pickups. In other words, neighboring intersections around shopping areas will exhibit similar (homogeneous) mobility patterns, likely corresponding to similar life-style behaviors, while the intersections around residential areas will exhibit different, yet still homogeneous mobility patterns and life-style behaviors; so, shopping areas will be very busy during the regular business hours and after work, but perhaps less so late in the evening, while residential areas will be busy on weekends and at times parents drop and pick up their kids from school. Similarly, in social networks, within a given social circle, users' profiles tend to be homogeneous, while within a different social circle they will be different, yet still homogeneous; for example, friends from a high school in New York City will probably have similar taste in entertainment, while friends from a high school in Lausanne will also have similar tastes but different from those from teenagers in New York.  We can model such data  as piecewise-smooth graph signals (see for example how a piecewise-smooth signal in Figure~\ref{fig:toy_pc}(b) provides a good approximation to the actual measurements of Figure~\ref{fig:toy_pc}(a)), as they capture large variations between pieces and small variations within pieces. 


\NEW{A number of communities consider similar questions as GSP. 
In particular, the machine learning community also considers graph structure/data  and at times,  uses similar methods as GSP does. For example, the graph Fourier basis is related to Laplacian eigenvectors and graph signal recovery is related to semi-supervised learning with graphs\cite{zhu2003semi}. There are, however, a few differences. (1) GSP defines a framework that allows the extension of classical signal processing concepts; (2) Sampling data associated with graphs is rarely studied in the machine learning community and the sampling problem on graphs is generally a hard one; graph-structured data representations (such as graph filter banks and graph dictionary learning) are also rarely studied in the machine learning community; (3) Given that GSP extends classical signal processing concepts, it is in a position to consider low-level processing such as denoising, inpainting and compression; and (4) Machine learning typically considers a graph as a discrete version of a manifold. In many real-world applications that are associated with graphs, this assumption is however not true. On the contrary, GSP does not make this assumption.
For example, there is no underlying manifold for online social networks. 
}
\NEW{In summary, as will be illustrated throughout this paper, GSP has strong connections to several theoretical and practical research domains; its promise lies in its ability to develop new tools and approach existing problems from different perspectives. }


\NEW{
\subsection{Related work in other areas}
\label{subsubsec:connectionsotherareas}
}

\NEW{
\subsubsection{Network science} This area addresses issues such as uncovering community relations, perceived alliances, quantifying connectedness, or determining the relevance of specific agents \cite{networksciencenas,borner2007network,lewis2011network,barabasi2016network}. Thus, much of this work does not concentrate on the data but rather its structure. \NEW{It determines for example} the size of the giant component, distribution of component sizes, degree and clique distributions, clustering coefficients, betweeness and closeness centralities, path length, and network diameter \cite{newman2010networks,jackson2010social,easley2010networks}. Connections to GSP are primarily due to graph spectra that GSP builds upon, \NEW{which is} strongly related to the structure of the graph \cite{Chung:96}. As an example, spectral clustering methods use the low frequency eigenvectors of the Laplacian \cite{von2007tutorial} and can thus be \NEW{addressed} from a GSP perspective \NEW{too} \cite{Tremblay:2016vb}. 
\subsubsection{Network processes} The aim is to model propagation over networks, including such phenomena  as diffusion of diseases and epidemics, spread of (fake) news, memes, fads, voting trends, imitation and social influence, propagation of failures and blackouts.  Common models are similar to stochastic automata where the states of the nodes (the ``data'') evolve through local rules, i.e., according to exogeneous (external to the network) and endogeneous (internal to the network) effects. For example, using terminology from epidemics, nodes of the graph representing agents or individuals of a population can be infected (adopt an opinion or spread a rumor), or susceptible (open to adopt an opinion or spread a rumor). Infected nodes can heal and become susceptible again; susceptible nodes can become infected either by an action external to the network or by action of infected neighbors \cite{castellano2012competing,nowzari2016analysis}. As the analysis of such  network processes is difficult, traditionally, the network is abstracted out, assuming that any node can infect any other node (full mixing or complete network).  To account for the impact of the network \cite{ganesh2005effect}, resorting to numerical studies is precluded except for very small networks since the network state space $\{0,1\}^N$ grows exponentially fast ($2^N$, for $N$ agents). To study these processes \cite{butts2009revisiting,colizza2006role} one usually considers one of two asymptotic regimes: (1) long term behavior (time-asymptotics),  attempting to find the equilibrium distribution of the process \cite{zhang2014diffusion,zhang2015role}; or (2) large network asymptotics (mean-field approximation) \cite{draief2010epidemics} leading to the study of the qualitative behavior of nonlinear ordinary differential equations \cite{santos2015bi}. Because asymptotic behavior can be seen to depend on the eigenstructure of the underlying graph, GSP representations as those discussed in Section~\ref{sec:representations} can be used to characterize the evolution of a system. As an example, several papers have explored the use of GSP techniques to improve the efficiency of value function estimation in a reinforcement learning scenario \cite{levorato2012structure,levorato2012reduced}. 
\subsubsection{Graphical models} The focus in this area is on inference and learning from large datasets, \cite{lauritzen1996graphical,jordan1998learning,wainwright2008graphical,koller2009probabilistic,whittaker2009graphical}. The data is modeled as a set of random variables described by a family of Gibbs probability distributions, and the underlying graph \NEW{(whose nodes label the variables)} captures statistical dependence and conditional independence among the data. Acyclic graphs \cite{bang2008digraphs,edwards2012introduction} represent Bayesian networks, and undirected graphs represent Markov random fields \cite{rozanov1982markov,kindermann1980markov,rue2005gaussian}. Graphical models exploit factorizations of the joint distribution to develop efficient message passing algorithms for inference and find application in many areas such as modeling texture and other features in image processing \cite{besag1974spatial,chellappa1993markov,moura1992recursive,balram1993noncausal,willsky2002multiresolution,vats2012finding}, see \cite{jordan2010major} for illustrative applications in several domains. Recent work on learning graph from data \NEW{\cite{Dong:2016fm,egilmez2017graph}}, which makes use of Markov random field models to define optimality criteria for the learned graphs, connects graphical models to GSP. 
\subsection{Historical perspective on graph signal processing}
\label{subsubsec:historicalperspective}
We now briefly review some of the prior work that is more directly connected and in the spirit of signal processing on graphs, \cite{SandryhailaM:13,ShumanNFOV:13}.  We organize the discussion along two main lines; \NEW{some parts of the exposition follow closely} \cite{SandryhailaM:13,Sandryhaila2014big}.
\subsubsection{From algebraic signal processing to graph signal processing} The sequence of papers \cite{Pueschel:03a,Pueschel:05e,Pueschel:08a,Pueschel:08b,Pueschel:08c} introduced algebraic signal processing (ASP), an axiomatic approach to time signal processing. ASP starts from a signal model~$\Omega$. Many signal models are possible, and a relevant question is to determine which one is more appropriate for a given application or should be associated with a given linear transform. Under appropriate conditions,  the signal model is generated from a simple filter, the \textit{shift}, which then determines filtering, convolution, the Fourier transform, frequency, and spectral analysis among other common concepts, and constructs from traditional digital signal processing. 
\NEW{Such formalism allowed for a uniform framework with variations of classical signal processing.}
ASP, after appropriately defining a \textit{space} line-graph signal model \cite{Pueschel:08b}, can be used to show that the DCT 
plays the same role for that signal model as the one the DFT plays for the time (cyclic) model.
ASP led to the introduction of, possibly weighted, graph adjacency matrices as shifts that generate the graph signal model for signals indexed by nodes of an arbitrary directed or undirected graph \cite{SandryhailaM:13,Sandryhaila:2014ju}. This choice is satisfying in the sense that, when the signal model is the classical time signal model, the shift and the graph signal model revert to the classical time shift (delay) and signal model~\cite{Pueschel:08a} (see Section~\ref{sec:asp-1}). Subsequently, authors have proposed other shifts obtained from the adjacency matrix of the 
graph~\cite{giraultgoncalvesfleury-2015,gavilizhang-2017} that attempt to preserve isometry of the shift, but in some cases lose the locality of the adjacency matrix shift \cite{giraultgoncalvesfleury-2015}
\subsubsection{From graph Laplacian spectral clustering to Laplacian-based GSP}
References \cite{Tenenbaum:00,Roweis:00,Belkin:03,Donoho:03} develop low-dimensional representations for large high-dimensional
data through spectral graph theory~\cite{Belkin:02,Belkin:03} and the graph Laplacian~\cite{Chung:96}, by projecting the data on a low-dimensional subspace generated by a small subset of the 
Laplacian eigenbasis \cite{von2007tutorial}. The use of the graph Laplacian is justified by assuming the data is smooth on the data space (manifold). References \cite{Coifman:05a,Coifman:05b,Coifman:06} choose discrete approximations to other continuous operators, for example, a conjugate to an elliptic Schr{\"o}dinger-type operator, and obtain other spectral bases for the characterization of the geometry of the manifold underlying the data.
}

\NEW{
Coming from another angle, motivated by processing data collected by sensor networks where sensors are irregularly placed, different authors develop regression algorithms \cite{guestrinbodikthibauxpaskinmadden-ipsn2004}, wavelet decompositions \cite{ganesangreensteinestrinheidemanngovindan-2005,wagnerbaraniuketal-SSPWorkshop2005,wagnerbaraniuketal-IPSN06, Coifman:06,Hammond:11}, filter banks on graphs \cite{Narang:10,Narang:12},  de-noising \cite{wagnerdelouillebaraniuk-2006}, and compression schemes using the graph Laplacian \cite{Zhu:12}.
Some of these references consider distributed processing of data from sensor fields, while others study localized processing of signals on graphs in a multiresolution fashion by representing data using wavelet-like bases with varying ``smoothness'' or defining transforms based on node neighborhoods. For example \cite{Hammond:11} uses the graph Laplacian and its eigenbasis to define a spectrum and a Fourier transform of a signal on a graph. 
Besides using the graph Laplacian, these works apply to data indexed by undirected graphs with real, non-negative edge weights. This approach is more fully developed in~\cite{ShumanNFOV:13}, which  adopts the graph Laplacian as basic building block to develop GSP for data supported by undirected graphs.
} 



\NEW{\subsubsection{Image processing, computer graphics and GSP}In addition, graph-based approaches have been widely used in signal processing contexts. For example, several authors representing images as graphs for segmentation \cite{wu1993optimal,shi2000normalized} and popular image-dependent filtering methods can be interpreted from a graph perspective \cite{milanfar2013tour}. Models used in computer graphics applications can often be viewed as graphs (e.g., meshes where vertices form triangles to which attributes are associated) and graph based filtering, processing and multi-resolution representations \cite{guskov1999multiresolution,zhou20043d,taubin1995signal}.}


\subsection{Outline of the paper}
The outline of the paper is as follows: Section~\ref{sec:asp-1} starts by presenting the framework and key ingredients of graph signal processing. It explains how the concepts from classical signal processing such as signals, filters and Fourier transform, among others, extend to complex structures where data is indexed by nodes on a graph. Section~\ref{sec:soa} covers some state-of-the-art topics and associated challenges, such as the definition of frequency, graph learning, sampling representations and others. Section~\ref{sec:applications} follows up with applications of graph signal processing in sensor networks, biological networks, 3D point cloud processing and machine learning. 
Section~\ref{sec:conclusion} gives some conclusions.


\section{Key Ingredients of Graph Signal Processing}
\label{sec:asp-1}

In this section we introduce basic graph signal processing~(GSP) concepts.
While more formal derivations of GSP can be developed, e.g., from the signal model introduced in the Algebraic Signal Processing~(ASP) \cite{Pueschel:03a,Pueschel:05e,Pueschel:08a,Pueschel:08b} or from the spectral perspective developed in \cite{ShumanNFOV:13,Hammond:11} based on spectral graph theory \cite{Chung:96}, we choose a more intuitive presentation by 
first reviewing the concept of shift in digital signal processing (DSP) (Section~\ref{sec:shift-DSP}) \NEW{in order to emphasize connections between DSP and GSP. We then develop} \sout{and then developing} a corresponding notion of shift for GSP (Section~\ref{sec:shift-GSP}). This in turns leads to the definition of frequencies for graph signals (Section~\ref{sec:frequency-GSP}) and their interpretation (Section~\ref{sec:Frequency-interpretation}). 
\NEW{
We focus on tools derived from the adjacency or Laplacian matrices of the graphs, as these are by far the most widely used. However, we note that each of these approaches have their own limitations and there are active research efforts to build GSP tools on alternative definitions of frequency (see Section~\ref{sec:frequency}).}

\subsection{The role of shifts in digital signal processing}
\label{sec:shift-DSP}

Discrete signal processing~(DSP) \cite{oppenheimschaffer-1975,oppenheimwillsky-1983,siebert-1986,oppenheimschaffer-1989,mitra-1998} studies time signals. Graph signal processing~(GSP)\footnote{We consider here only linear graph signal processing.} \cite{SandryhailaM:13,ShumanNFOV:13,Sandryhaila2014big} extends DSP to signal samples indexed by nodes of a graph. 
At a very high level, DSP, and therefore GSP, study:
\begin{inparaenum}[1)]
\item signals and their representations;
\item systems that process signals, usually referred to as filters;
\item signal transforms, including two very important ones, namely, the $z$-transform and the Fourier transform; and
\item sampling of signals, as well as other more specialized topics.
\end{inparaenum}

Consider $N$ samples of a signal~$s_n$, $n=0,1,\cdots,N-1$. We restrict ourselves to signals with a finite number~$N$ of samples and to filters with finite impulse response (FIR filters). 
The $z$-transform~$s(z)$ of the (real or complex valued) time signal $s=\left\{s_n:n=0,1,\cdots,N-1\right\}$ organizes its samples~$s_n$ into an ordered set of time samples, where sample~$s_n$ at time~$n$ precedes~$s_{n+1}$ at time~$n+1$ and succeeds~$s_{n-1}$ at time~$n-1$. In other words, the signal is given by the $N$-tuple $s=\left(s_0, s_1,\cdots, s_{N-1}\right)$.  This representation is achieved by using a formal variable, say~$z^{-1}$, called the shift (or delay), so that the signal of $N$-samples is represented by
\begin{align}
\label{eqn:ztransform-1}
s(z)=\sum_{n=0}^{N-1}\,s_nz^{-n}.
\end{align}
The $z$-transform $s(z)$ provides a (formal)\footnote{\NEW{While in DSP $z$ is a complex variable, which leads to the DFT when restricted to the unit circle as in (\ref{eqn:signalDFT1}), here we establish the link to GSP by viewing $z$ as a placeholder for each sample of the signal.}} polynomial representation of the signal that is useful in studying how signals are processed by filters. Clearly, given $s(z)$ we can recover the signal~$s$ \cite{oppenheimschaffer-1989,mitra-1998}.

The discrete Fourier transform (DFT) of the signal~$s$ is~$\widehat{s}=\left\{\widehat{s}_k:\,k=0,\cdots,N-1\right\}$ given by
\begin{align}
\label{eqn:signalDFT1}
\widehat{s}_k=\frac{1}{\sqrt{N}}\sum_{n=0}^{N-1}\,s_ne^{-j\frac{2\pi}{N}kn}.
\end{align}
The $\widehat{s}_k$ are the Fourier coefficients of the signal. The DFT represents the signal~$s$ in the dual or frequency domain, leading to concepts such as frequency, spectrum, low-, band-, and high-pass signals. 
The discrete frequencies are 
\NEW{$\Omega_k= \frac{2\pi k}{N}$, }
$k=0,1,\cdots, N-1$, and the $N$~signals $\left(x_k[n]\right)$
\begin{align*}
\left\{x_k[n]=\frac{1}{\sqrt{N}}e^{-j\frac{2\pi}{N}kn}:\:n=0,1,\cdots, N-1\right\}_{k=0}^{N-1}
\end{align*}
 are the spectral components.

 The signal is recovered from its Fourier coefficients by the inverse DFT: 
\begin{align}
\label{eqn:signalinverseDFT1}
s_n=\frac{1}{\sqrt{N}}\sum_{k=0}^{N-1}\,\widehat{s}_ke^{j\frac{2\pi}{N}kn}, s=0,1,\cdots, N-1.
\end{align}

In DSP, besides signals, we also have filters~$h$. An FIR filter is also represented by a polynomial in $z^{-1}$
\begin{align}
\label{eqn:filterhz}
h(z)=\sum_{n=0}^{N-1} h_nz^{-n}, 
\end{align}
so that the output~$s_{\scriptsize\textrm{out}}$ of filter~$h$ applied to signal~$s_{\scriptsize\textrm{in}}$ is:
\begin{align}
\label{eqn:filter_output}
s_{\scriptsize\textrm{out}}(z)=h(z)\cdot s_{\scriptsize\textrm{in}}(z).
\end{align}
Because we are only considering finite time signals, and the product above could result in $s_{\scriptsize\textrm{out}}(z)$ being a polynomial in $z^{-1}$ of degree greater than $N-1$, we have to consider boundary conditions (b.c.). For simplicity, we consider periodic extensions of the signal, i.e., the signal sample $s_N$ is equal to the signal sample $s_0$; more generally, $s_n=s_{n\!\!\!\mod\!\!N}$. In other words, the real line is folded around the circle. 
%
Defining the shift or delay filter
\begin{align*}
h_{\scriptsize\textrm{shift}}(z)=z^{-1},
\end{align*}
and applying it to a signal $s_{\scriptsize\textrm{in}}=\left(s_0,s_1,\cdots,s_{N-1}\right)$ gives an output: 
\begin{align*}
s_{\scriptsize\textrm{out}}=h_{\scriptsize\textrm{shift}}\cdot s_{\scriptsize\textrm{in}}=\left(s_{N-1},s_0, s_1,\cdots,s_{N-2}\right).
\end{align*}
By Equation~\eqref{eqn:filterhz}, any filter~$h$ in DSP is a polynomial in the shift, i.e., it is built from series and parallel combinations of shifts. Thus, the shift is the basic building block in DSP, from which we can build more complicated filters.


A second very important DSP property that is adopted in GSP is \textit{shift invariance}. This readily follows from
\begin{align}
\label{eqn:DSPshiftinvariance}
z^{-1}\cdot h(z)=h(z)\cdot z^{-1}.
\end{align}
In words, the series combination of filters is commutative, a filter commutes with the shift filter\textemdash delaying the input signal $s_{\scriptsize\textrm{in}}$ and then filtering the delayed input signal leads to the same signal as first filtering the input signal $s_{\scriptsize\textrm{in}}$ and then delaying the filtered output.

\begin{cframed}[red]
Restating for emphasis, both~\eqref{eqn:ztransform-1} and~\eqref{eqn:filterhz} show the principal role played by the shift~$z^{-1}$ in DSP. We represent signals by (finite degree) polynomials in $z^{-1}$ and build filters also as polynomials in $z^{-1}$.
\end{cframed}

\subsection{Defining shifts in Graph Signal Processing} 
\label{sec:shift-GSP}

We now extend the above concepts and tools to \textit{graph} signals, i.e., signals whose samples are indexed by the nodes of arbitrary graphs. 
To do so, we start by reinterpreting the finite signals from the previous section as vectors rather than tuples or sequences.

Rewrite the signal $s=\left(s_0, s_1,\cdots,s_{N-1}\right)$ as the vector
\begin{align*}
\bm{s}=\left[s_0\,s_1\,\cdots\,s_{N-1}\right]^\top\,\in\mathbb{C}^N,
\end{align*}
where for generality we allow the signal to be complex valued. Using this notation, a filter~$h$ is represented by a matrix~$\Hm$ and   (\ref{eqn:filter_output}) can be  simply written as a matrix-vector multiplication:
\begin{align*}
\bm{s}_{\scriptsize\textrm{out}}=\Hm\cdot \bm{s}_{\scriptsize\textrm{in}},
\end{align*}
where filters are represented by matrices, while signals are represented by vectors. In particular, the shift filtering operation corresponds to multiplication by a \NEW{circulant} matrix~$\Am_c$
\begin{align*}
\left[s_{N-1}\,s_0\,\cdots\,s_{N-2}\right]^\top= \Am_c\cdot \left[s_0\,s_1\,\cdots\,s_{N-1}\right]^\top,
\end{align*}
given by the cyclic shift
\begin{align}
\label{eqn:Ashiftcyclic}
\Am_c=\left[\begin{array}{cccccc}
0&0&0&\cdots&0&1\\
1&0&0&\cdots&0&0\\
0&1&0&\cdots&0&0\\
\vdots&\vdots&\ddots&\ddots&\ddots&0\\
0&0&\cdots&1&0&0\\
0&0&\cdots&0&1&0
\end{array}\right].
\end{align}

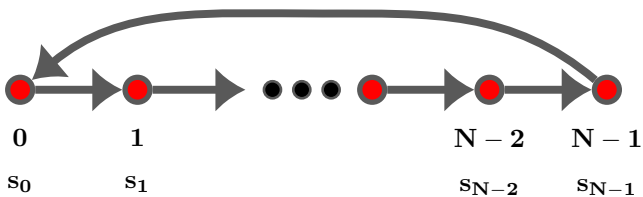
\begin{figure}[htb]		
        \centering
\begin{tikzpicture}[scale=0.78]
	\node[draw=gray!70!black,circle,fill=red,line width=2pt] (0) at (0,0) {};
	\node[draw=gray!70!black,circle,fill=red,line width=2pt] (1) at (2,0) {};
	\node (2) at (4,0) {};
	\node[draw=gray!70!black,circle,fill=red,line width=2pt] (N3) at (6,0) {};
	\node[draw=gray!70!black,circle,fill=red,line width=2pt] (N2) at (8,0) {};
	\node[draw=gray!70!black,circle,fill=red,line width=2pt] (N1) at (10,0) {};
	
	\draw[-{Latex[length=4mm,width=6mm]},line width=3pt,draw=gray!70!black] (0) -- (1);
    \draw[-{Latex[length=4mm,width=6mm]},line width=3pt,draw=gray!70!black] (1)-- (2);
	\node[draw=gray!70!black,circle,fill=black,line width=1pt,inner sep=.8mm] (dot1) at (4.3,0) {};
	\node[draw=gray!70!black,circle,fill=black,line width=1pt,inner sep=.8mm] (dot1) at (4.8,0) {};
	\node[draw=gray!70!black,circle,fill=black,line width=1pt,inner sep=.8mm] (dot1) at (5.3,0) {};
	\draw[-{Latex[length=4mm,width=6mm]},line width=3pt,draw=gray!70!black] (N3) -- (N2);
	\draw[-{Latex[length=4mm,width=6mm]},line width=3pt,draw=gray!70!black] (N2) -- (N1);

	\node (tmp) at (5,1.3) {};
	\draw[-{Latex[length=4mm,width=6mm]},line width=3pt,draw=gray!70!black] (N1) to[out=140,in=0] (tmp) to[out=180,in=40] (0);	

	\node[below = 2mm of 0] (0_id) {\bf $\mathbf{0}$};
	\node[below = 2mm of 1] (1_id) {\bf $\mathbf{1}$};
	\node[below = 2mm of N2] (N2_id) {\bf $\mathbf{N-2}$};
	\node[below = 2mm of N1] (N1_id) {\bf $\mathbf{N-1}$};

	\node[below = 1.5mm of 0_id] (0_sig) {\bf $\mathbf{s_0}$};
	\node[below = 1.5mm of 1_id] (1_sig) {\bf $\mathbf{s_1}$};
	\node[below = 1.5mm of N2_id] (N2_sig) {\bf $\mathbf{s_{N-2}}$};
	\node[below = 1.5mm of N1_id] (N1_sig) {\bf $\mathbf{s_{N-1}}$};
    

    
\end{tikzpicture}
\caption{Time graph: Cycle graph $G_c$.\label{fig:cyclegraph}}
\end{figure}

A graph interpretation for the DSP concepts of Section~\ref{sec:shift-DSP} can be achieved by viewing the 0-1 shift matrix $\Am_c$ of~\eqref{eqn:Ashiftcyclic} as the adjacency matrix of a graph. Labeling the rows and columns of~$\Am_c$ from~0 to~$N-1$, define the graph $G_c=(V,E)$ with node set $V=\left\{0,1,\cdots, N-1\right\}$. 
\NEW{Row~$n$ of~$\Am_c$ represents the set of in-edges of node~$n$ in~$G_c$\textemdash if there is an entry~1 at column~$\ell$, $A_{c,n\ell}=1$, then there is an edge from~$\ell$ to~$n$. $\Am_c$ is then the adjacency matrix of the cycle graph in Figure~\ref{fig:cyclegraph}.}

\begin{cframed}[blue]
The key point we make is the dual role of the matrix~$\Am_c$ in Equation~\eqref{eqn:Ashiftcyclic}, which represents both the shift~$z^{-1}$ in DSP and the adjacency matrix of the associated time graph in Figure~\ref{fig:cyclegraph}.
\end{cframed}


This graph interpretation of DSP can be extended to develop a linear time shift invariant Graph Signal Processing \cite{SandryhailaM:13}. 
Consider now a graph signal $\bm{s}\in\mathbb{C}^N$, where the entries of the signal~$\bm{s}$ are indexed by the $N$~nodes of an arbitrary graph $G=(V,E)$, $v_1,...,v_N$. Assuming that the graph has edge weights $w_{ij}$, denote an edge of weight $w_{ij}$ going from $v_j$ to $v_i$, then we can define the following algebraic representations associated to $G$. 
\NEW{\begin{definition}[Algebraic representations of graphs]
\label{def:useful_matrices}
The \textbf{adjacency matrix} is a matrix, $\Am$, such that $(\Am)_{ij} = w_{ij}$. \\
In the particular case where the graph is undirected, we have $w_{ij}=w_{ji}$, \NEW{$\Am$ is now symmetric,} and we also define 
the \textbf{degree matrix} of $G$, a diagonal matrix, $\mathbf{D}$, with entries $(\mathbf{D})_{ii}= \sum_{j=1}^{N} (\Am)_{ij}$ and $(\mathbf{D})_{ij} = 0$ for $i \neq j$,  
the \textbf{combinatorial graph Laplacian} defined as $\Lm \!=\!\mathbf{D}\!-\!\Am$, and the {\bf symmetric normalized Laplacian} ${\mathbf{\cal L}} = \Dm^{-1/2}\Lm \Dm^{-1/2} $. 
\end{definition}}

The adjacency matrix~$\Am$ can be adopted as the shift~\cite{SandryhailaM:13} for this general graph. Other choices have been proposed, including the Laplacians \cite{ShumanNFOV:13}, or variations of these matrices \cite{giraultgoncalvesfleury-2015,gavilizhang-2017}. Different choices for the shift present different trade-offs. The adjacency matrix~$\Am$ reduces to the shift in classical time DSP and applies to directed and undirected 
graphs\footnote{\NEW{Note that the graph defined by $\Am_c$ in (\ref{eqn:Ashiftcyclic}) is directed in order to match exactly the behavior of shifts in time in DSP, which are always directed, i.e., we either move forward or backwards in time. But in general the notion of a graph shift applies to any adjacency matrix, whether corresponding to a directed or an undirected graph. In what follows both directed and undirected graphs are considered.} }, while the graph Laplacian applies only to undirected graphs, \NEW{so that}~$\Lm$ is symmetric and positive semi-definite, which avoids a certain number of analytical and numerical difficulties that may arise when choosing~$\Am$. Furthermore, graph Laplacian spectra have been widely studied 
in the field of spectral graph theory \cite{Chung:96}. In specific applications, one should consider 
definitions and choose the one that leads to 
the best trade-off for the problem being considered \cite{chen2015signal-2}. \NEW{This choice is further discussed in Sections~\ref{sec:selection} and~\ref{sec:frequency}}. 


For time signals, as discussed with respect to~\eqref{eqn:ztransform-1}, the basis $\left\{z^{-n}\right\}_{n=0}^{N-1}$ orders the samples of the signal by increasing order of the time labels (nodes in time graph). Rewriting~\eqref{eqn:ztransform-1}, we get
\begin{align*}
s(z){}&=\left[\left(z^{-1}\right)^0\,z^{-1}\cdots z^{-(N-1)}\right]\left[s_0\,s_1\cdots s_{N-1}\right]^\top.
\end{align*}
In graph signal processing, ordering the samples corresponds to labeling the nodes of the graph. This labeling or numbering fixes the adjacency matrix of the graph, and hence the graph shift. The columns of the graph shift provide a basis and a representation for the graph signals. Other bases could be used, leading to different signal representations.
We note that relabeling the nodes of the graph by a permutation~$\Pim$ conjugates the shift by~$\Pim$
\begin{align*}
\Am_{\Pi}=\Pim \Am\Pim^\top.
\end{align*}


Following the analogy with DSP, we can now define the notion of {\bf shift invariance} and {\bf polynomial filters} for arbitrary graphs. 
A filter represented by~$\Hm$ will be shift invariant
if it commutes with the shift,
\begin{align*}
\Am\Hm=\Hm\Am.
\end{align*}
As proven in \cite{SandryhailaM:13}, if the characteristic polynomial $p_A(z)$ and 
the minimum polynomial\footnote{\NEW{For a matrix $\Am$ the minimal polynomial $m_A(z)$ is the polynomial of minimal degree having $\Am$ as a root. }}
$m_A(z)$ of~$\Am$ are equal, then every filter commuting with~$\Am$ is a polynomial in~$\Am$, i.e.,
\begin{align*}
\Hm=h(\Am).
\end{align*}
For equality $p_A(z)=m_A(z)$, to each eigenvalue of~$\Am$ there corresponds a single eigenvector\footnote{In other words, the Jordan form of~$\Am$ has single blocks for each distinct eigenvalue.}. A simpler condition is for the eigenvalues of~$\Am$ to be distinct. To keep the discussion simple, unless otherwise stated, we assume~$\Am$ has~$N$ distinct eigenvalues and hence a complete set of eigenvectors. 

By the Cayley-Hamilton Theorem of Linear Algebra \cite{gantmacher1959matrix,lancaster1985theory}
\begin{align*}
\textrm{degree}(h(z))=\textrm{degree}(p_A(z))\leq N-1.
\end{align*}
In fact, $\textrm{degree}(h(z))\leq \textrm{degree}(m_A(z))\leq \textrm{degree}(p_A(z))$. In words, shift invariant filters are polynomials with degree at most $\textrm{degree}(m_A(z))$.

\subsection{Frequency representations for graph signals}
\label{sec:frequency-GSP}

In DSP and in linear systems, we are interested in signals that are invariant when processed by a (linear) filter, i.e., 
\begin{align*}
h\cdot s_{\scriptsize{\textrm{in}}}=\alpha s_{\scriptsize{\textrm{in}}},
\end{align*}
where $\alpha$ is a scalar (from the base field). Such $s_{\scriptsize{\textrm{in}}}$ are of course the eigensignals of the filter~$h$. In GSP we define filters as matrices and thus the eigensignals of~$h$ are the eigenvectors of the corresponding~$\Hm$. More interestingly, since shift invariant filters are polynomials of a single matrix, the shift~$\Am$, we only need to consider the eigenvectors of~$\Am$. Then, write 
\begin{align}
\label{eqn:diagonalizationA}
\Am=\Vm\Lam\Vm^{-1} 
\end{align}
where~$\Vm=\left[\bm{v}_0\cdots\bm{v}_{N-1}\right]$ is the matrix of the~$N$ eigenvectors of~$\Am$, $\Lam=\textrm{diag}\left[\lambda_0\cdots \lambda_{N-1}\right]$ is the matrix of distinct eigenvalues of~$\Am$. \NEW{Because we assume~$\Am$ has a complete set of eigenvectors, $\Vm$ is invertible. }
%
%
Then, it is straightforward to verify that for each (polynomial) filter
\begin{align}
\nonumber
\Hm{}&=h(\Am)\\
\nonumber
{}&=h\left(\Vm\Am\Vm^{-1}\right)\\
\nonumber
{}&=\sum_{m=0}^{M-1}h_m \left(\Vm\Lam\Vm^{-1}\right)^m\\
\label{eqn:fiterresponse}
{}&=\Vm h\left(\Lam\right)\Vm^{-1}, 
\end{align}
where $h\left(\Lam\right)$ is the diagonal matrix
\begin{align}
\label{eqn:filterfrequencyresponse2}
h\left(\Lam\right)=\textrm{diag}\,\left[h\left(\lambda_0\right)\cdots h\left(\lambda_{N-1}\right)\right].
\end{align}
We can promptly verify that the eigenvectors of~$\Am$ are the eigenfunctions of the (polynomial) filter
\begin{align}
\nonumber
\Hm \bm{v}_m{}&=\Vm h\left(\Lam\right)\Vm^{-1}\bm{v}_m\\
\nonumber
{}&=\Vm h\left(\Lam\right)\bm{e}_m\\
\label{eqn:filtereigenresponse1}
{}&=h\left(\lambda_m\right)\bm{v}_m,
\end{align}
where $\bm{e}_m$ is the zero vector except for entry~$m$ that is a one.  Equation~(\ref{eqn:filtereigenresponse1}) is the GSP counterpart to the classical DSP fact that exponentials are eigenfunctions of linear systems. As such the response of the filter to an exponential is the same exponential amplified or attenuated by a gain that is the frequency response of the filter at the frequency of the exponential. We refer to this as the \textit{invariance} property of exponentials with respect to linear systems in DSP. Accordingly,  Equation~\eqref{eqn:filtereigenresponse1} shows the invariance of the eigenvectors of the shift operator~$\Am$ with respect to graph filters.

Finally, we can introduce the Fourier transform for graph signals. 
The cyclic shift in Equation~\eqref{eqn:Ashiftcyclic} can be written as 
\begin{align}
\label{eqn:DFT1}
\Am_c={\bf DFT}_{N}^{-1}\left(\begin{matrix}e^{-j\frac{2\pi\cdot 0}{ N}}&&\cr&\ddots&\cr&&e^{-j\frac{2\pi\cdot(N-1)}{ N}}\end{matrix}\right){\bf DFT}_{N},
\end{align}
where ${\bf DFT}_{N}=\frac{1}{\sqrt{N}}\left[\omega_N^{kn}\right]$, $\omega_N=\exp^{-j\frac{2\pi}{N}}$, is the discrete Fourier matrix. The inverse ${\bf DFT}_{N}^{-1}={\bf DFT}_{N}^H$ is the matrix of eigenvectors of~$\Am_c$. The eigenvalues of~$\Am_c$ are $e^{-j\frac{2\pi\cdot n}{ N}}$, $n=0,\cdots,N-1$. the diagonal entries of the middle matrix in Equation~\eqref{eqn:DFT1}.
The graph Fourier transform (GFT) follows by analogy with~\eqref{eqn:DFT1}. From the eigendecomposition of~$\Am$ in~\eqref{eqn:diagonalizationA}, the graph Fourier transform is the inverse of the matrix~$\Vm$ of eigenvectors of the shift~$\Am$
\begin{align}
\label{eqn:graphFouruier1}
\Fm=\Vm^{-1}. 
\end{align}
The eigenvectors of the shift~$\Am$, columns of~$\Vm$, are the graph spectral components, and the eigenvalues of~$\Am$, the diagonal entries $\lambda_k$ of matrix~$\Lam$ in~\eqref{eqn:diagonalizationA}, are the graph frequencies. The graph frequencies are complex valued for a general non-symmetric (directed graph) shift~$\Am$.

The graph Fourier transform of graph signal~$\bm{s}$ is given by the graph Fourier \textit{analysis} decomposition
\begin{align}
\label{eqn:fourieranalysis1}
\widehat{\bm{s}}=\Fm\bm{s}=\Vm^{-1} \bm{s}=\left[f_0\bm{s}\cdots f_{N-1}\bm{s}\right]^\top, 
\end{align}
\NEW{where $f_k$ is \NEW{a row vector,} the $k$-th row of $\Fm$. The graph Fourier coefficients or graph spectral coefficients of signal~$\bm{s}$ are computed using the inner product as $\widehat{s}(\lambda_k) = \widehat{s}_k=f_k \bm{s} = \left\langle f_k^{H},\bm{s} \right\rangle$.
Then, the Fourier spectral decomposition of the signal is obtained by the graph inverse Fourier transform. Equivalently, it is given by the graph Fourier synthesis expression
\begin{align}
\nonumber
\bm{s}{}&=\Fm^{-1}\widehat{\bm{s}}=\Vm \widehat{\bm{s}}\\
\label{eqn:fouriersynthesis1}
{}&=\sum_{k=0}^{N-1}\,\widehat{s}_k v_k\\
\nonumber
{}&=\sum_{k=0}^{N-1}\,\left\langle f_k^{H},\bm{s} \right\rangle v_k\\
\nonumber
{}&=\Vm\left[\left\langle f_0^{H},\bm{s} \right\rangle\cdots \left\langle f_{N-1}^{H},\bm{s} \right\rangle\right]^\top.
\end{align}
}
The eigenvectors $v_k$ of~$\Am$, columns of $\Vm$, are the spectral components. Equation~\eqref{eqn:fouriersynthesis1} synthesizes the original signal~$\bm{s}$  from the spectral components $v_k$; the coefficients $\widehat{s}_k$ of the decomposition are the spectral coefficients of~$\bm{s}$. 

\subsection{Interpreting Graph Frequencies}
\label{sec:Frequency-interpretation}


We can now interpret filtering a graph signal (i.e., multiplying the corresponding vector by $\Hm$) in the spectral domain. From~\eqref{eqn:fiterresponse}, the output of $\bf{s}_{\scriptsize\textrm{in}}$ to filter~$h$ is successively
\begin{align}
\nonumber
\bm{s}_{\scriptsize\textrm{out}}{}&=\Hm\cdot\bm{s}_{\scriptsize\textrm{in}}\\
\label{eqn:firstFourier1}
{}&=\Vm h\left(\Lam \right)\underbrace{\left(\Vm^{-1} \bm{s}_{\scriptsize\textrm{in}}\right)}_{\scriptsize\textrm{Fourier transf.}}\\
\nonumber
{}&=\Vm\underbrace{\textrm{diag}\left[h\left(\lambda_0\right)\cdots h\left(\lambda_{N-1}\right)\right] \widehat{\bm{s}}_{\scriptsize\textrm{in}}}_{\scriptsize\textrm{Filtering in graph Fourier space}}\\
\label{eqn:pointwisemult-1}
{}&=\underbrace{\Vm\left[h\left(\lambda_0\right)\widehat{\bm{s}}_{{\scriptsize\textrm{in}}_0}\cdots h\left(\lambda_{N-1}\right)\widehat{\bm{s}}_{{\scriptsize\textrm{in}}_{N-1}}\right]^\top}_{\scriptsize\textrm{Inverse Fourier transf.}}.
\end{align}
Thus, according to ~\eqref{eqn:firstFourier1}, filtering by $\Hm$ can be performed by first taking the graph Fourier transform of the input $\left(\Vm^{-1}\bm{s}_{\scriptsize\textrm{in}}\right)$, followed by pointwise multiplication in the frequency domain of the graph Fourier transform signal $\widehat{\bm{s}}_{\scriptsize\textrm{in}}$ by the \textit{filter frequency response} $\left[h\left(\lambda_0\right)\cdots h\left(\lambda_{N-1}\right)\right]^{\top}$ given  by~\eqref{eqn:pointwisemult-1}. Finally, an inverse graph Fourier transform computes the output back in the graph node domain. This is the graph Fourier filtering Theorem that reduces graph filtering to two graph Fourier transforms  and a pointwise multiplication in the spectral domain
\cite{SandryhailaM:13}. 

With a notion of frequency we can now consider the GSP equivalents to classical concepts of low-, high-, and band-pass signals or filters, as well as the question of efficient filter design. In the classical time domain, these concepts are directly related to values of \textit{frequency}. 
\NEW{In the time domain, the}
frequency is actually defined from the eigenvalues of the cyclic shift~$\Am_c$ in~\eqref{eqn:DFT1} as
\begin{align*}
\NEW{\Omega_k= \frac{2\pi k}{N} , \: k=0,1,\cdots,N-1.}
\end{align*}
These frequencies are directly related to the degree of variation of the  spectral components. For example the lowest frequency $\Omega_0=0$ corresponds to the least varying spectral component, the constant or DC-spectral component, 
\NEW{the next frequency $\Omega_1=\frac{2\pi}{N}$ represents a higher variation spectral component, and so on. }
There is a nice one-to-one correspondence between the ordered value of the frequency and the corresponding degree of variation or \textit{complexity} of the time spectral component.

\begin{figure*}[h]
\centering
    \includegraphics[width=0.95\linewidth]{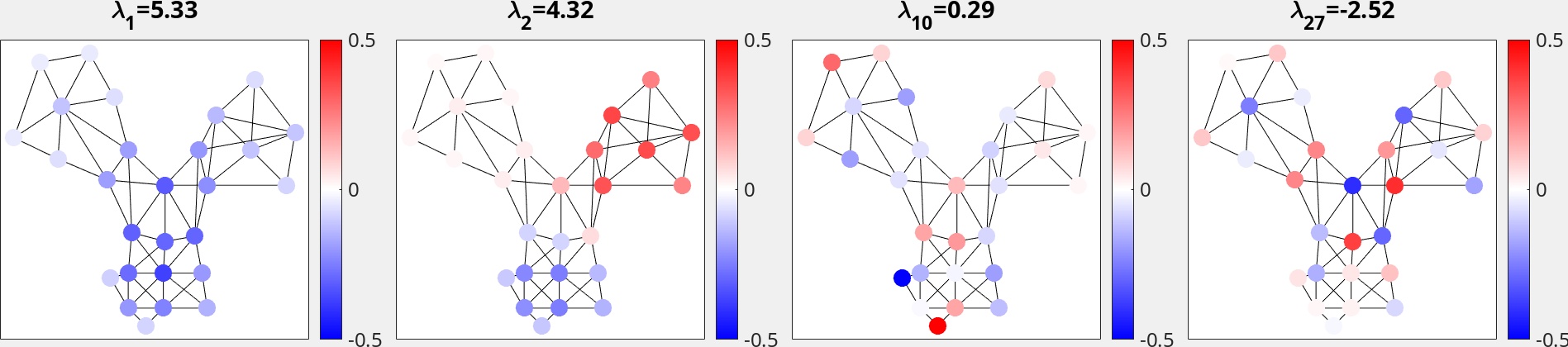}\par\medskip
    \mbox{\centering{\small(a)}}\par\medskip
    \includegraphics[width=0.95\linewidth]{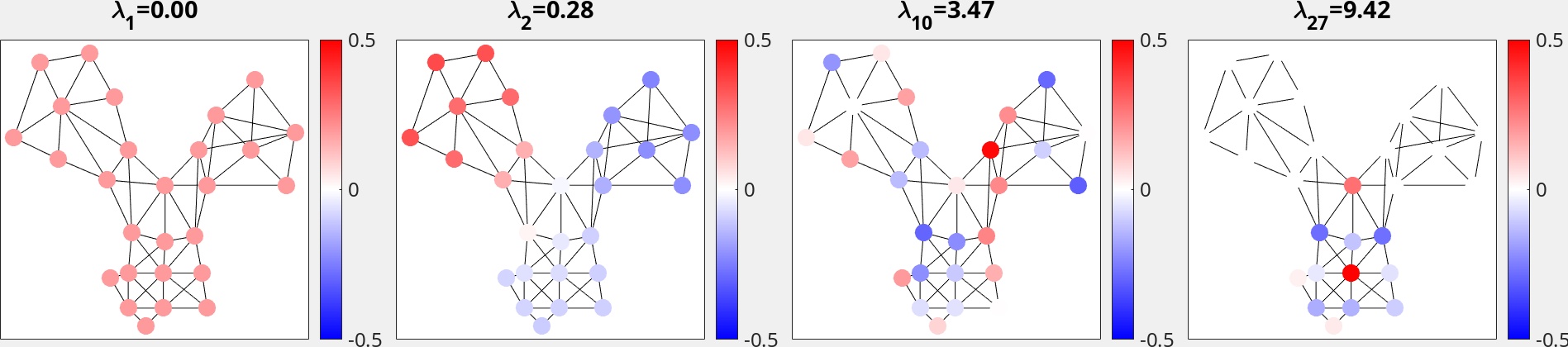}\par\medskip
     \mbox{\centering{\small(b)}}
\caption{\NEW{\label{fig:eigenvector_example} Example of elementary frequencies obtained from different algebraic representations of the same graph. (a) Adjacency \NEW{matrix} (b) Laplacian matrix. In each case 4 different frequencies are shown, corresponding to different eigenvalues, ranging from lowest frequency to highest frequency. In the Laplacian case the lowest frequency is $\lambda=0$, representing a constant value throughout the graph, the highest is $\lambda=4.53$, where we can observe a large number of sign changes across graph edges. Note that for any given graph with $N$ nodes we will have $N$ eigenvectors that can be \NEW{ordered} in terms of their variation covering the whole range of frequencies for that graph. In this example the graph is unweighted. 
\NEW{Unlike} conventional signal processing some of the eigenvectors can be localized in the graph (e.g., the highest frequency eigenvector of the Laplacian).}}
\end{figure*}

In GSP, the frequencies are defined by the eigenvalues of the shift. 
We can order the graph frequencies by 
relating them to the \textit{complexity} of the spectral component.  
For example, this can be measured by the \textit{total variation} of the associated spectral component through
\begin{align*}
\textrm{TV}_{\scriptsize\textrm{G}}\left(\bm{v}_k\right)= \left\|\bm{v}_{k}-\Am^{\scriptsize\textrm{norm}}\bm{v}_{k}\right\|_1,
\end{align*} 
where $\|\cdot\|_1$ is norm~1, and $\Am^{\scriptsize\textrm{norm}}=\frac{1}{\lambda_{\scriptsize\max}}\Am$. Other norms could be used to define the total variation, see \cite{Sandryhaila:2014ju}\cite{ShumanNFOV:13}. Using this, graph frequency $\lambda_m$ is larger than graph frequency $\lambda_\ell$ if
\begin{align*}
\textrm{TV}_{\scriptsize\textrm{G}}\left(\bm{v}_m\right)> \textrm{TV}_{\scriptsize\textrm{G}}\left(\bm{v}_\ell\right).
\end{align*}
Assuming the graph frequencies have been ordered from low to high, graph signal~$\bm{s}$ is low-pass if its graph Fourier coefficients are zero for $\Omega_k$, $k>\ell$, \NEW{for some $\ell$,} $0\leq \ell<N-1$. We can similarly define band- and high-pass signals and filters\footnote{\NEW{The total variation is the $l1$ norm of a vector multiplied by $\Id - \Am^{\scriptsize\textrm{norm}}$. Assume for simplicity that the graph is undirected, then the largest eigenvalue of 
$\Id - \Am^{\scriptsize\textrm{norm}}$, and thus the largest TV, will be $1-\lambda_{min}/\lambda_{max}$, where $\lambda_{min}$ is the smallest  eigenvalue of $\Am$, which intuitively, as seen in Fig.~\ref{fig:eigenvector_example}, corresponds to high variation in the eigenvector. }}.

\NEW{\subsection{Frequency representations based on the Laplacian}
\label{sec:selection}}

The notions of frequency that arise in conventional signal processing provide a sound mathematical and intuitive basis for analyzing signals. While it is mathematically possible, as just discussed, to define  notions of frequency for graph signals, developing a corresponding intuition to understand these elementary frequencies is not as straightforward. 
For the total variation criterion it has been shown experimentally and justified theoretically that the frequency bases obtained from the shift operator tend to be ordered \cite{Sandryhaila:2014ju}. 

\NEW{Up to this point, we have focused primarily on frequency representations derived from the adjacency matrix of a graph, an approach that can be applied to both directed and undirected graphs, and can be linked to DSP concepts in the case of the cycle graph. 
A frequency representation can be similarly built on top of the Laplacian matrix of an undirected graph. Since this matrix is positive semidefinite, all the eigenvalues are real and non-negative, and a full set of orthogonal eigenvectors can be obtained, so that we can write  
\begin{equation}
\Lm = \Um{\bf \Lambda}\Um^{\top}
\end{equation}
with $\Um$ the  GFT matrix, which is real valued and orthogonal in this case. Because the eigenvalues are real, they provide a natural way to order the GFT basis vectors in terms of frequency \NEW{(the variations of their values on the graph).} In this case, the eigenvalue/eigenvector pairs can be viewed as successive optimizers of the Rayleigh quotient, where the $k$-th pair, $\lambda_k, \uv_k$ solves :
\begin{equation}
\uv_k = \arg\min_{\xv^{\top}\uv_{k'} = 0,\, k'=0, \, \ldots, k-1}\frac{\xv^{\top}\Lm\xv}{\xv^{\top}\xv}
\label{eq:rayleigh}
\end{equation}
with $\lambda_k = \uv_{k}^{\top}\Lm\uv_{\uv_k}$, if $\uv_k$ is normalized. Thus for the explicit variation metric induced by the Laplacian quadratic form, the GFT provides an orthogonal basis with increased variation, and such that, from (\ref{eq:rayleigh}), each additional basis vector minimizes the increase in variation while guaranteeing orthogonality.}
More generally, the relationships between eigenvectors and eigenvalues of the Laplacian and the structure of a graph are part of a deep and beautiful domain of mathematics known as spectral graph theory~\cite{Chung:96}. When graphs have structures closely related to those used in DSP (e.g., circulant Adjacency matrices \NEW{\cite{ekambaram2015spline}}) frequency interpretation is clear. If the graph is more general than the ring graph, part of the intuition remains, as illustrated by Figure~\ref{fig:eigenvector_example}. 
Indeed, eigenvectors $u_i$ are oscillating over the vertex set. As the eigenvalue index $i$ increases, \NEW{the number of oscillations tends to increase as well~\cite{briandavies2001discrete}.} 
However, the irregular nature of graphs means that the analogies to DSP cannot always be extended easily. For example, the spacing between frequencies (as measured by the eigenvalues of the Laplacian, for example) can be highly irregular, or some frequencies may have high multiplicity. \NEW{Also, the high frequency eigenvectors of irregular graphs can be highly 
localized \cite{brooks2013non,saito2011phase}. This potentially indicates that a direct ordering of frequencies may be insufficient to fully understand signal decompositions induced by current GSP techniques. \NEW{To be complete, note that,} while Laplacians can be easily defined for undirected graphs, there has been work to introduce definitions appropriate for directed graphs as well \cite{chung2005laplacians,bauer2012normalized}.}
\NEW{In summary, a full understanding of the best frequency representation for a specific GSP application, as a function of the type of graph considered, is still an active research topic. 
This is discussed further in Section~\ref{sec:frequency}.}

\subsection{Implementation}

Finally, let us quickly touch on the issue of computational complexity of the filtering operation. 
A straightforward algorithm would consist in computing the GFT matrix $\Vm$ and explicitly applying it to the input signal as in (\ref{eqn:firstFourier1}). This is simple and accurate for small graphs thanks to fast SVD algorithms. Partial SVD can also be used if the filter $h$ should only be evaluated on the top or bottom eigenvalues \NEW{\cite{golub2012matrix}}. In general, and for large graphs, it is better to avoid computing even a partial SVD. One efficient possibility is to compute a polynomial approximation to $h$ with Chebyshev filters~\cite{Hammond:11}. For large but sparse graphs, this reduces computations to sparse matrix-vector multiply, which is very efficient.

\NEW{Furthermore, filter implementation via polynomial approximation can be interpreted in terms of localization in the vertex domain.  }
Note that when the input signal is a perfect impulse located at a given vertex, $\sv = \ev_i$, the filtered signal depends only on the graph filter and the vertex location in the graph: $\fv_i = \Hm \ev_i$. Even though $\fv_i$ changes with the chosen vertex, it was proved in \cite{Localization} that this signal is localized around $i$ in a way that only depends on the smoothness of the filter $h$. This is interesting because it allows to design filters that act locally and in a controlled way over the vertex set. \NEW{After a filter $h(\lambda)$ is chosen, one can choose an approximation $h_k(\lambda)$, a polynomial of degree $k$ in $\lambda$. Note that $h_k(\lambda)$ can then be implemented as shown in (\ref{eqn:fiterresponse}) by applying a polynomial of the shift operator. This does not require knowing the eigenvalues and eigenvectors associated to the graph, so that it is possible to process signals on very large graphs locally, by processing $k$-hop neighborhoods of nodes in the vertex domain, without a need to find the graph spectrum first.}



\section{State-of-the-Art and Challenges}
\label{sec:soa}

\subsection{Frequency definition}
\label{sec:frequency}

One can guarantee the existence of an orthogonal basis for any undirected graph. Thus, once a graph has been chosen (see Section~\ref{sec:graph-learning}) a definition of frequency is readily available, which allows us to address other questions considered in this section (sampling, signal representation, etc). Multiple choices are possible, as a function of the graph type, the selected shift operator and its normalization, etc. Making these choices appropriately for a given application remains an open question, which is actively being investigated. 

As an example, the eigenvalues of the chosen operator matrix (Laplacian or adjacency) can have high multiplicity. In this situation, a graph with $N$ nodes will have fewer than $N$ unique frequencies. 
A particular concern is that one can choose any set of orthogonal vectors within the subspace corresponding to this frequency, leading to different GFTs and thus potentially irreproducible results. As a way to address this scenario, recent work \cite{deri2017spectral} suggests using oblique projections to measure the energy within such a subspace, using this information to represent the overall energy at that frequency. 

For directed graphs, additional problems arise given that a full set of eigenvectors may not exist. Results for directed graphs are often restricted to cases where the adjacency matrix is invertible and eigenvectors do exist (as discussed in Section~\ref{sec:frequency-GSP}). If these conditions do not hold, the Jordan canonical form is used to obtain the GFT \cite{SandryhailaM:13}, but this is well known to be a numerically unstable procedure. 
As an alternative, some authors have proposed to approximate directed graphs by undirected ones, using such approaches as the hub-authority model \cite{kleinberg1999authoritative,zhou2005semi}. \NEW{ Recent work has also considered alternative definitions of frequency. For example, the work in \cite{mhaskar2016unified} advocates using  the random walk Laplacian normalization, while in \cite{girault2018irregularity} the authors propose alternative choices of a graph signal inner product and explore the resulting frequency definitions. Other techniques make use of explicit optimization to choose a set of graph frequencies. }
As an example, the work in \cite{sardellitti2017graph} uses an optimization procedure to construct explicitly an orthogonal basis set that minimizes a quantity related to the cut size. With this approach, successive  eigenvectors provide increasingly higher frequencies in the sense of corresponding to higher cut costs, while being orthogonal to those eigenvectors previously selected. \NEW{The work in \cite{shafipour2017digraph} also uses optimization techniques with a different criterion to define a set of frequencies associated to a graph. } 
In summary, this is a very active area of research, and the best approach to define a set of frequencies for graphs in a specific application remains to some extent an open question. 

\subsection{Representations}
\label{sec:representations}

Designing representations for graph signals having desirable properties (e.g., localization, critical sampling, orthogonality, etc) has been one of the first and most important research goals in graph signal processing. Pioneering contributions \cite{Crovella2003} and \cite{Coifman:06}, provided early examples of designs based on vertex domain and spectral domain characteristics, respectively. Vertex domain designs such as \cite{Crovella2003} or \cite{narang2009lifting} have the advantage of defining exactly localized basis functions on the graph, but do not have a clear spectral interpretation. 
Conversely, diffusion wavelets \cite{Coifman:06} are defined in the spectral domain, but do not guarantee exact vertex domain localization (only energy decay properties). The spectral graph wavelet transform design \cite{Hammond:11} was the first to combine a spectral design with vertex-domain localization, by defining smooth filter kernels in the spectral domain and approximating these with polynomials. 

The filterbanks developed in \cite{Hammond:11} were not critically sampled, unlike \cite{Coifman:06} or \cite{narang2009lifting}. Thus, much recent work has focused on developing critically sampled filterbanks having both a spectral interpretation and vertex localized implementation. These types of filterbanks have been designed for bipartite graphs \cite{Narang:12,narang2013compact}, thus requiring the graph to be decomposed into a series of bipartite subgraphs \cite{Narang:12,zeng2017bipartite}.
\NEW{An alternative approach proposed in \cite{kotzagiannidis2017splines,ekambaram2015spline} can be applied to circulant graphs, for which the GFT corresponds to the DFT.}
Recent work \cite{teke2017extending1,teke2017extending2} has shown that similar filterbank designs can be developed for directed graphs, where these designs are only possible for $M$-block cyclic graphs, which play a similar role to that of bipartite graphs in the undirected case. \NEW{Note that in all these cases, critical sampling combined with polynomial analysis and synthesis filtering is restricted to specific types of graphs (bipartite, $M$-block cyclic and circulant.) Note also that critical sampling with polynomial analysis and synthesis filters on undirected graphs can only be achieved in the bipartite case \cite{anis2017critical}\footnote{Under some conditions on the analysis filters, critical sampling and perfect reconstruction can be achieved for any graph, but this requires a synthesis operation corresponding to an $N \times N$ matrix multiplication, which may not be practical for large graphs \cite{anis2017critical}. As an example, the approach in \cite{ekambaram2015spline} guarantees invertibility but reconstruction is non polynomial.}. } Ongoing work is focusing on i) providing better tools to characterize $M$-block cyclic graphs, including for example the definition of polyphase representations \cite{teke2017extending1,teke2017extending2,tay2015techniques,tay2017bipartite}, ii) development of improved filters by exploiting conventional filter designs and/or relaxing the critical sampling requirement \cite{tanaka2014m,sakiyama2014oversampled,tay2015design,tay2017critically}, \NEW{ and iii) novel approaches for downsampling, e.g., frequency domain techniques \cite{tanaka2017spectral}, that allow extending critically sampled filterbanks to non-bipartite graphs.}

While much of the work to date has focused on representations with bases functions selected in terms of frequency content (e.g., low pass vs.~high pass bases), some recent work is also exploring representations for piecewise smooth signal models \cite{ChenSK:17}. The design of representations that adapt to the specific properties of graph signal classes has further been addressed from the viewpoint of dictionary learning \cite{Zhang:2012fu,Thanou:2014gj,Yankelevsky:2016gka}. The main objective is to design dictionary of atoms that are able to sparsely represent signals on graphs while incorporating the structure of the graph. 


\subsection{Sampling}
\label{sec:sampling}

The problem of sampling signals on graphs is modeled on the corresponding problem in conventional signal processing. The basic idea is to define a class of signals (for example signals that are  bandlimited to the first $K$ frequencies of the GFT) and then define necessary and sufficient conditions to reconstruct a signal in that class from its samples.  
The first problem formulation and a sufficient condition for unique recovery were presented  in~\cite{pesenson2008sampling}. A necessary and sufficient condition for unique recovery in undirected graphs was introduced in \cite{anis2014towards}, and subsequently several papers proposed solutions for different aspects of the problem \cite{shomorony2014sampling,ChenVSK:15,anis2016efficient}. In particular, sampling results have been generalized to directed graphs \cite{ChenVSK:15,anis2016efficient} and to other classes of signals such as piecewise smooth signals \cite{chen2015signal}. 

A key difference when comparing sampling in conventional signal processing and in the context of graph signals is the lack of ``regular'' sampling patterns in the latter. The lack of regularity in the graph itself prevents us from defining the idea of sampling ``every other node''. Thus, multiple approaches have been suggested to identify the most informative vertices on a graph so that these can be sampled. While the sampling problem is formalized based on the assumption that signals to be sampled belong to a certain class (e.g., bandlimited), in practice these can never be guaranteed and thus the observed signals will be noisy and in general will not belong to the pre-specified class. To address this problem, several methods approach the problem of sampling set selection from an experiment design perspective \cite{gadde2015probabilistic,ChenVSK:15,anis2016efficient} setting as a goal to identify a set of vertices that minimizes some measure of worst case reconstruction error in cases where noise or model mismatch is present. The measure can also be mean squared reconstruction error instead of worst case in the experiment design paradigm \cite{anis2016efficient}. 

Complexity is a key challenge in sampling set identification, especially for large-scale graphs. Some techniques require computing and storing the first $K$ basis vectors of the GFT \cite{ChenVSK:15}. For larger graph sizes, where this may not be practical, the approach in \cite{anis2016efficient} uses spectral proxies instead of exact graph frequencies leading to lower complexity. To reduce complexity even further, the work in \cite{puy2016random} proposes a random sampling technique where the probability of selecting a given vertex is based on a locally computed metric. This leads to significantly lower complexity but, as a random sampling technique, it may not always lead to performance comparable to those of more complex greedy optimization methods such as \cite{ChenVSK:15,anis2016efficient}. 

Given the samples of a graph signal, the next objective is to reconstruct an estimation of the signal at the nodes that were not sampled (observed). 
Reconstruction algorithms based on polynomial filters approximating ideal reconstruction filters have been proposed in order to reconstruct an estimated signal on the whole graph based on the observed vertex measurements \cite{narang2013signal,wang2015local}. 

While theoretical aspects of graph signal sampling are by now well understood, the relevance of proposed techniques to practical applications is still an open question. A key challenge in this regard is to identify what are relevant signal models for real datasets, while potentially adapting proposed generic sampling methods to specific types of graphs (e.g., exploiting properties of nearly regular graphs).


\subsection{Extending conventional signal processing to graph signals}

Challenges in extending ideas and concepts from conventional signal processing to signal processing on graphs can be further exemplified by research into notions of stationarity and localization. For conventional time signals, a test for stationarity can be based on determining whether time shifts affect the statistical properties of a signal or, equivalently, observing a signal at different times. However, these two views are not equivalent for finite dimension graphs: we can observe a given signal at different nodes, but this is not necessarily the same as ``shifting'' the signal while  observing it at always at the same node.  For graphs with $N$ vertices, shifting can be defined via a spectral domain operator \cite{Hammond:11}; or, instead, the graph shift based on the adjacency matrix can be used. Some authors have proposed a definition of stationarity based on spectral properties of the vertex shift operator  \cite{marques2016stationary}. To overcome challenges associated to existing shift operators, one solution, first proposed by  \cite{girault2015stationary}, is to introduce alternative graph shift operators (see also  \cite{giraultgoncalvesfleury-2015}) or localization operators that have both a spectral interpretation and vertex domain localization \cite{girault2016localization,perraudin2017stationary}. \NEW{Notions of stationarity can help develop probabilistic graph signal processing \NEW{methods} leading to graph-based Wiener filtering \cite{girault2014semi,perraudin2017stationary}.}

A study of vertex/spectral localization and uncertainty principles was first developed by \cite{Agaskar:12}, where it was shown that  in general it is not possible to achieve arbitrarily good localization in both spectral and vertex domains simultaneously. However, a limitation in this study was that bounds had to be derived for individual vertices. More recently, \cite{tsitsvero2016signals} has shown that for graph signals it is in fact possible to have compact support in both spectral and vertex domain (something that can never occur in conventional signal processing). \NEW{As  was already noted in Section~\ref{sec:selection}}, this occurs due to the irregular nature of graphs: for example, a graph consisting of several loosely connected clusters is likely to lead to some columns of $\Vm$ having non-zero entries only in some of the clusters.  Other contributions, such as \cite{pasdeloup2015toward,teke2017uncertainty}, have also explored the challenges in directly extending the concept of an uncertainty principle to graph signals, \NEW{while other recent 
work considers alternative frequency representations that can take into consideration the specific localization properties encountered in irregular graphs \cite{behjat2016signal,van2017slepian,irion2014hierarchical} }

Work in these two areas shows that direct extensions of signal processing concepts to graphs are not straightforward, and thus further research is still needed to develop techniques that can provide insights about graph signal behavior (localization, stationarity) while accommodating key characteristics of graphs (e.g., irregular node connectivity and spectral characteristics).

\subsection{Graph learning}
\label{sec:graph-learning}

Much recent work on graph signal processing assumes that the graph is given or can be defined in a reasonable way based on the nature of the application. As an example, in communication or social networks the connectivity of the network (directed or undirected) can be used to define the graph. In other applications edge weights between nodes can be chosen as a decreasing function of distance, e.g., physical distance between sensors in the case of sensor networks or distance in feature space in the case of learning applications \cite{zhu2003semi,jebara2009graph,gadde2014active}. 


Recent work has been considering alternative techniques where the goal is to learn graphs from data. This work is motivated by scenarios where i) no reasonable initial graph exists or ii) it is desirable to modify a known graph (based on network connectivity for example) by selecting weights derived from data. The key idea in these approaches is to select a graph such that the most likely vectors in the data (the graph signals) correspond to the lowest frequencies of the GFT or to the more likely signals generated by Gauss Markov Random Field (GMRF) related to the graph. 

Examples of approaches based on smoothness include \cite{Dong:2016fm,Kalofolias:2016tf,daitch2009fitting}, while representative methods based on the GMRF model are  \cite{lake2010discovering,egilmez2017graph}. The basic idea in the latter approaches is to identify GMRF models such that the inverse covariance (precision) matrix has the form of a graph Laplacian (e.g., combinatorial or generalized). Note that this work extends popular approaches for graph learning (e.g., graphical Lasso \cite{friedman2008sparse}) to precision matrices restricted to have a Laplacian form (corresponding to a graph with positive edge weights). 
Other approaches have addressed graph selection under the assumption that the observed data was obtained through graph-based diffusion. Examples of these approaches include \cite{mei2017signal,pasdeloup2017characterization,Segarra_templates,Thanou:2017ec}. \NEW{While not explicitly a graph learning problem, the related question of blind identification of graph filters has also been studied \cite{segarra2017blind}.}

\NEW{There remain several major challenges in the development of graph learning methods. Graphs derived from data are essentially models, and as such the ``right'' graph model should be selected based on the number of parameters it uses, its data fit and its ability to provide useful interpretations. While a sparsity criterion addresses some of these requirements, other constraints may also be important. For example, it will be useful to develop methods to select graphs with specific topology properties~\cite{pavez2017learning}, spectral properties (eigenvalue distribution, eigenvector localization), or even computational properties (e.g., leading to GFTs with lower computation cost.)  }

\section{Graph Signal Processing Applications}
\label{sec:applications}

Networks are present in very different application domains, where graphs can provide a generic representation of the structure present in the datasets. In this section, we discuss a wide set of applications where the graph signal processing framework has been used. We consider four different types of scenarios, where both the scale and the domain of the networks considered are very different. We start with physical networks, including both large scale networks (sensor networks in Section~\ref{sec:sensors}) and human-scale ones (biological networks in Section~\ref{sec:biological}), where the goal is to use measurements to better understand physical phenomena. We then consider ``logical'' networks, where GSP is introduced as an alternative for existing processing techniques for conventional signals (images and point clouds in Section~\ref{sec:images}), or as a tool to analyze large scale datasets (machine learning and data science applications in Section~\ref{sec:machine-learning}). 
In each of these cases we provide a few, non-exhaustive, examples to highlight the different types of domains and graph representations that have been studied. More detailed discussion of graph-based techniques in specific domains are considered in other papers in this special issue \cite{cheung2018graph}.


\subsection{Sensor networks}
\label{sec:sensors}

One of the most natural applications of Graph signal processing is in the context of sensor networks. A graph represents the relative positions of sensors in the environment, and the application goals include compression, denoising, reconstruction, or distributed processing of sensor data. Indeed, some of the initial explorations of graph-based processing focused on sensor networks \cite{wagnerbaraniuketal-SSPWorkshop2005,wagnerbaraniuketal-IPSN06,ciancio2006energy,shen2008joint}. 

A first approach to define a graph associated to a sensor network is to choose edge weights as a decreasing function of distance between nodes (sensors). Then, data observations that are similar at neighboring nodes lead naturally to a smooth (low-pass) graph signal. Such a smooth graph signal model makes it possible to detect outliers or abnormal values by high-pass filtering and thresholding \cite{Sandryhaila:2014ju,egilmez2014spectral}, or to build effective signal reconstruction methods from sparse set of sensor readings, as in \cite{Zhu:2012wc,Kaneko:2017ui,sakiyama2016efficient}, which can potentially lead to significant savings in energy resources, bandwidth and latency in sensor network applications. 

A second scenario is where the graph to be used for data analysis is given by the application. For example, urban data processing  relies on data that naturally live on networks, such as energy, transportation or road networks. In these applications cases, GSP has been used to monitor urban air pollution \cite{Jain:db}, or to monitor and analyze power consumption \cite{He:2016jg}, for example. Some works such as \cite{Valdivia:2015ji,Chen:2016tt,Dong:2013eg} have used graph signal processing tools for analyzing traffic and mobility in large cities. For example, wavelets on graphs can serve to extract useful traffic patterns to detect disruptive traffic events such as congestion \cite{Mohan:2014cs}. Graph wavelet coefficients at different scales permit to infer useful information such as origin, propagation, and the span of traffic congestion. 

In some cases, relations between sensor readings are not exclusively explained by the distance between sensor locations, or by some actual network constraints. Other factors can influence the data values observed at the sensor readings such as the presence of geographical obstacles (e.g., in temperature measurements), or the interaction between networks of different types (e.g., how proximity to a freeway affects pollution measurements in a city). In some cases the phenomena that can explain these relations between measurements are latent and this leads to the challenging problem of learning a graph (see also Section~\ref{sec:graph-learning}) that can explain the data observations under signal smoothness or other signal model assumptions \cite{Dong:2016fm,Kalofolias:2016tf,Thanou:2017ec}. This allows inferring  system features and behaviors that are hidden in the measured datasets (e.g., ozone datasets in \cite{Jablonski:js}). 

Finally, several of the graph signal processing operators presented in this paper are amenable to distributed implementations that are particularly interesting for large sensor networks, and which motivated some of the early work mentioned at the beginning of this section. For example, the graph multiplier operators can be approximated by Chebyshev polynomials in distributed implementation of smoothing, denoising, inverse filtering or semi-supervised learning tasks \cite{Shuman:2017ti}. The work in \cite{Wang:2015ik} for example studies the problem of distributed reconstruction of time-varying band-limited graph signals recorded by a subset of temperature sensor nodes. There is however still a lot of opportunities for the development of distributed GSP algorithms that are able to extend to large-scale networks and big data applications. 

\subsection{Biological networks}
\label{sec:biological}

\begin{figure*}[t!]		
        \centering
        \includegraphics[trim=1cm 0cm 0cm 0cm, clip=true,width=0.98 \textwidth]{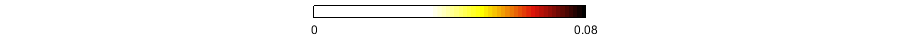}
	
        \centering
        \includegraphics[width=0.98 \textwidth]{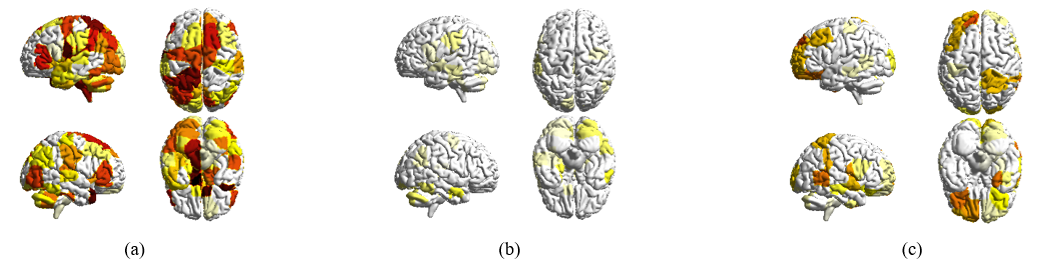}
        \caption{Distribution of decomposed signals. (a), (b), and (c) are the absolute magnitudes for all brain regions with respect to graph low frequencies, graph middle frequencies and graph high frequencies, respectively. Higher concentration in graph low frequency results in better learning performance, when subjects are unfamiliar with the task \cite{huang2016}. Concentration in graph low frequency also helps faster response in switching attention between actions \cite{medaglia2018}. From \cite{huang2016}, with permission.
}
        \label{fig_brain_signal_analytics}
\end{figure*}
 
Biological networks have also proved to be a popular application domain for graph signal processing, with recent research works focusing on the analysis of data from systems known to have a network structure, such as 
the human brain, and also on the inference of a priori unknown biological networks.
  
Several works have studied human brain networks using the graph signal processing framework. For example, it has been observed that human brain activity signals can be mapped on a network (graph) where each node corresponds to a brain region. The network links (edge weights) are considered to be known a priori and represent the structural connectivity or the functional coherence between brain regions \cite{bullmore2012, sporns2011}. GSP tools such as the graph signal representations described in Section~\ref{sec:representations} can then be used to analyze the brain activity signal on the functional or structural brain network. For example, low frequencies in the graph signal represent similar activities in regions that are highly connected in the functional brain networks, while high frequencies denote very different activities in such brain regions. 

These ideas have been to analyze brain signals, leading to biologically plausible observations about the behavior the human cognitive system, as in for example \cite{huang2016, medaglia2018}. Fig. \ref{fig_brain_signal_analytics} illustrates the signal distribution of different graph frequency components in an active motor learning task. Interestingly, regions with strong signal in low and high graph frequency components overlap well with the regions known to contribute to better motor learning \cite{Bassett2011}. Additionally, it has been observed that there is a strong association between the actual brain networks (characterized by their spectral properties) and the level of exposure of subject to different tasks \cite{Goldsberry:2017ha}. Some works further build on the multi-resolution properties of spectral graph wavelet transforms to capture subtle connected patterns of brain activity or provide biologically meaningful decompositions of functional magnetic resonance imaging (fMRI) data \cite{Behjat:2015gp,Leonardi:2013bv,Atasoy:2016ev}. Interestingly, it is also possible to combine different sources of informations in the analysis of the brain networks. For example, the work in \cite{Griffa:2017in} integrates infra-slow neural oscillations and anatomical-connectivity maps derived from functional and diffusion MRI, in a multilayer-graph framework that captures transient networks of spatio-temporal connectivity. These networks group anatomically wired and temporary synchronized brain regions and encode the propagation of functional activity on the structural connectome, which contributes to a deeper understanding of the important structure-function relationships in the human brain.

The GSP framework has also been proposed for the classification of brain graph signals \cite{Menoret:2017usa} and the analysis of anomalies or diseases \cite{Hu:2015fe,Hu:2016fw}. For example, source localization algorithms based on sparse regularization can be used to localize the possible origins of Alzheimer's disease based on a large set of repeated magnetic resonance imaging (MRI) scans. This can help understand the dynamics and origin of dementia, which is an  important step towards developing effective treatment of neuro-degenerative diseases \cite{Hu:jn}. The growing number of  publications studying brain activity or brain network features from a GSP perspective indicates that these are promising applications for the methods described in this paper. 

It should finally be noted that brain networks are not the only biological networks where GSP offers promising solutions. Graph signal processing elements and biological priors are combined to infer networks and discover meaningful interactions in gene regulatory networks, as in  \cite{Pirayre:2015di,Pirayre:2017cm}. The inference of the structure of protein interaction networks has also been addressed with help of spectral graph templates \cite{Segarra_templates}. In particular, the observed matrix of mutual information can be approximated by some (unknown) analytic matrix function of the unobserved structure to be recovered. Observed data is then used to obtain the eigenvectors of the matrix representation of the graph model, and then the eigenvalues are estimated with the help of sparsity priors. The above examples are only some illustrations of the recent works that attempt to infer structures of biological networks using a GSP perspective. Biological networks that cannot be explicitly recorded and measured are potentially good applications for  graph learning and inference methods in particular, which can uncover unknown interactions in the biological data.   

\subsection{Image and 3D point cloud processing}
\label{sec:images}

While graph signal processing is often applied to datasets that naturally exhibit irregular structures, it has also been applied to other datasets where conventional signal processing has been used for many years, including for example images and video sequences. An image to be processed can be viewed as a set of pixels, each associated to a vertex, forming a regular graph with all edge weights equal to 1 (e.g., a line graph or a grid graph). Indeed processing using the discrete Fourier transform or the discrete cosine transform (DCT) can be shown to have a simple interpretation in terms of the frequencies associated to those regular graphs \cite{strang1999discrete}  
(see also Section~\ref{subsubsec:historicalperspective}). 
Instead, recent work uses regular line and grid graph topologies, but with unequal edge weights that can adapt to the specific characteristic of an image or a set of images. 

A first set of approaches  associates a different graph to each image, by associating smaller edge weights to connect pixels that are on opposite sides of an image contour. 
This type of image-dependent graph representation is strongly connected to popular image processing techniques, such as the bilateral filter and related methods \cite{milanfar2013tour}, which also apply signal dependent filtering and are widerly used in  applications such as image restoration or denoising. Graphs are used to capture the geometric structure in images, such as contours that carry crucial visual information, in order to avoid blurring them during the filtering process. In addition to works that effectively extend image priors such as Total Variation (TV) minimization to graph representations (e.g., \cite{Couprie:2013fd,Najman:2017ug}), other works such as \cite{Tian:hy} or \cite{Pang:2015jo} use more specific GSP operators for denoising or filtering. In particular, the authors in \cite{Tian:hy} use graph spectral denoising methods to enhance the quality of images, while the work in \cite{Pang:2015jo} uses graph-based filters that influence the strength and direction of filtering for effective enhancement of natural images.

A second avenue of research has considered situations where a graph is constructed as an efficient representation for a set of images, in particular in the context of image and video compression applications. The Karhunen-Lo\`eve transform (KLT) is known to provide the best transform coding gains under the assumption that the signals can be modeled as stationary Gaussian processes (which is often a good assumption for images). Indeed, extensive use of the DCT is often justified because it is optimal for a Gauss Markov Random Field (GMRF) with correlation $1$, which is an appropriate model for natural images. The inverse covariance matrix, or precision matrix, then corresponds to a line graph with equal weights. From this perspective, graph learning approaches can be used to learn precision matrices with structures and weights that capture statistics of specific types of images. For example, piecewise smooth images can be compressed using suitable Graph Fourier Transforms (GFT), which can be adapted to different types of image pixel blocks \cite{Hu:2015fs,Fracastoro:2016wn}. Graph-based transforms have also been used to code motion-compensated residuals in predictive video coding \cite{Egilmez:gh} with effective rate-distortion performance. 

  \begin{figure}[t]
 	\centering
		\includegraphics[width=7.2cm]{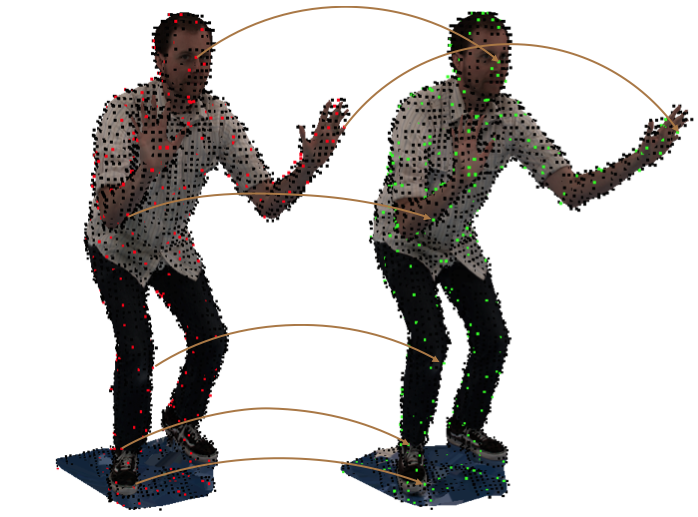}
           \caption{ Example of motion estimation in a 3D point cloud sequence. Each frame is represented as a graph signal that captures the color and the geometry information of each voxel. Graph spectral features at each voxel capture the local graph signal properties and permit to find correspondences between frames at different time instances. A subset of the correspondences between the target (red) and the reference frame (green) are highlighted between small cubes that correspond to voxels. From \cite{Thanou:di}, with permission.} 
           \label{fig:3dmotion}
\end{figure}

New visual modalities such as 3D meshes or 3D point clouds where data is sampled in irregular locations in 3D space,  lend themselves naturally to graph representations. The color or 3D information supported by nodes or voxels are connected to their nearest neighbors to form a graph. Graph-based transforms can then be used to compress the resulting graph signals in static or dynamic point clouds \cite{Zhang:2014ey,Thanou:di}. In particular, the temporal redundancy between 3D point cloud frames at different time instants can be effectively estimated with help of graph spectral features \cite{Thanou:di}, as illustrated in Figure \ref{fig:3dmotion}. Graph-based transforms permit to properly exploit both the spatial correlation inside each frame and the temporal correlation between the frames, which eventually results in effective compression. Compression, however, is not the only application of GSP in 3D point clouds. Fast resampling methods, which are important in processing, registering or visualizing large point clouds, can also be built on graph-based randomized strategies to select representative subsets of points while preserving application-dependent features \cite{ChenTFVK:16}. 

\subsection{Machine Learning and Data Science}
\label{sec:machine-learning}

Graph methods have long played an important role in machine learning applications, as they provide a natural way to represent the structure of a dataset. In this context, each vertex represents one data point to which a label can be associated, and a graph can be formed by connecting vertices with edge weights that are assigned based on a decreasing function of the distance between data points in the feature space. Graph signal processing then enables different types of processing, learning or filtering operations on values associated to graph vertices. In a different context, GSP elements can be helpful to construct architectures that are able to classify signals that live on irregular structures. We give below some examples of machine learning applications in both contexts.  

When data labels are presented as signals on a (nearest neighbor) graph, graph signal regularization techniques can  be used in the process of estimating labels \cite{zhu2003semi}, optimizing the prediction of unknown labels in classification \cite{Sandryhaila:2014ju} or semi-supervised learning problems \cite{ChenCRBGK:13}. Furthermore, as labeled samples are often a scarce and expensive resource in semi-supervised learning applications, graph sampling strategies such as those presented in Section~\ref{sec:sampling} can be helpful in determining the actual needs for labeled data and develop effective active learning algorithms \cite{gadde2014active}. 

Graphs can also be constructed to describe similarities between users or items in recommendation systems that assist customers in making decisions by collecting information about how other users rate particular services or items \cite{Benzi:fm}. Leveraging the notions of graph frequency and graph filters, classical collaborative filtering methods (such as $k$-nearest neighbors strategies), can then be implemented with specific \textit{band-stop} graph filters on graphs \cite{Huang:2017er}. Furthermore, linear latent factor models, such as low-rank matrix completion, can be viewed as \textit{bandlimited} interpolation algorithms that operate in a frequency domain given by the spectrum of a joint user and item network. This can serve to design effective graph filtering algorithms that lead to enhanced rating prediction in video recommendation applications, for example \cite{Huang:2017er}. Content-based recommendation can also be addressed as an online learning problem solved with spectral bandit algorithms \cite{Valko:2014wi}. The key idea is to represent the reward function in an online recommendation system as a linear combination of the eigenvectors of the similarity graph that connects the different items. With this representation it is possible to optimize the reward function by favoring smoothness on the graph, which has been shown to be effective in video recommendation examples \cite{Valko:2014wi}.  

\begin{figure*}[tb]
\begin{minipage}[b]{.99\linewidth}
\centering \includegraphics[width=0.3\textwidth]{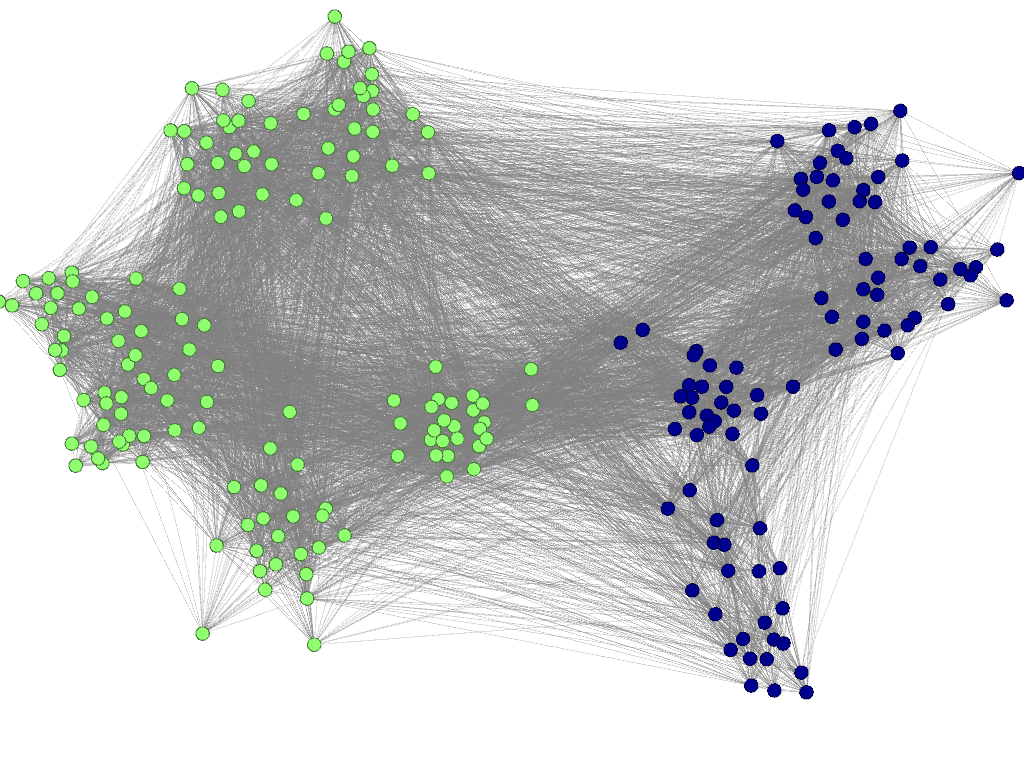} \hspace{0.2cm} \includegraphics[width=0.3\textwidth]{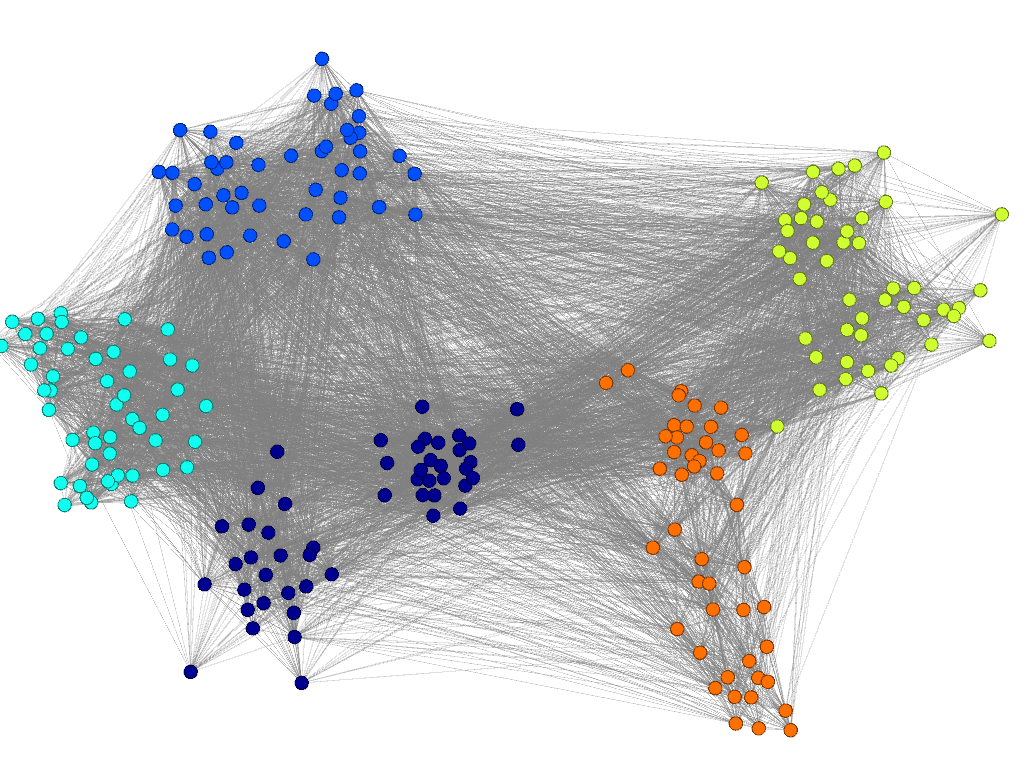} \hspace{0.2cm} \includegraphics[width=0.3\textwidth]{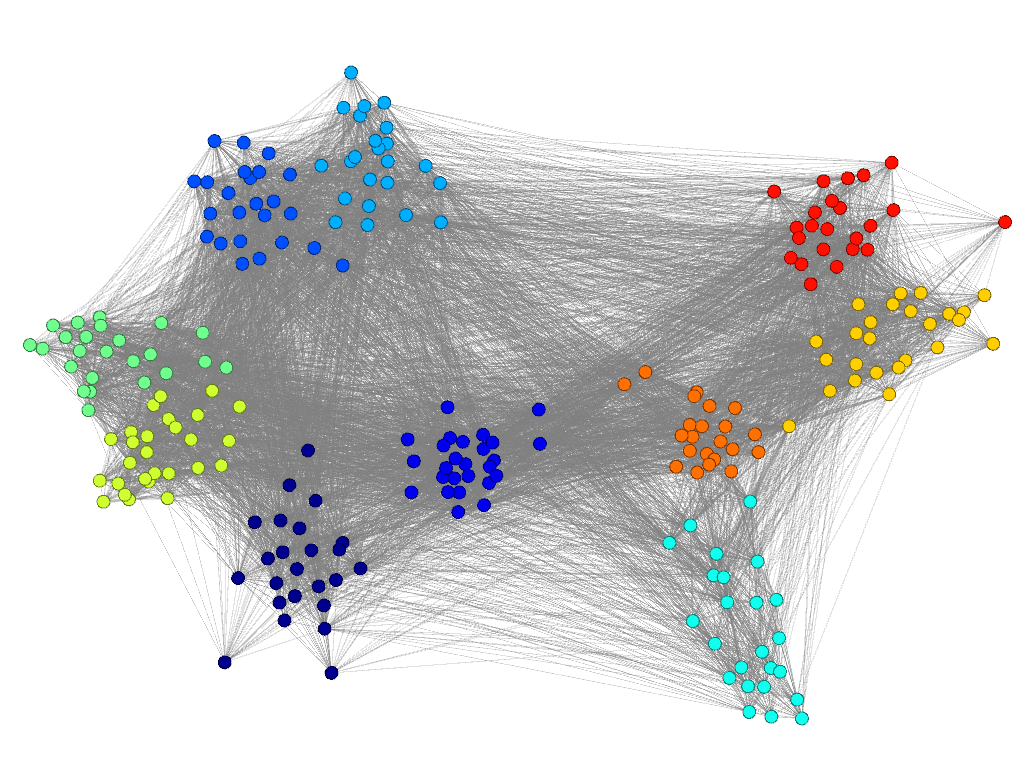}
\end{minipage}
\caption{Multiscale community structures in a graph of social interactions between children 
in a primary school. The different figures show the partition of the original social network in 2, 5 and 10 communities, respectively. From \cite{Tremblay:2014hl}, with permission.}
\label{fig:communities}
\end{figure*}

Data clustering or community detection can also benefit from  tools developed under the GSP framework. 
For example graph transforms, and especially graph wavelets, have been used to solve the classical problem of community detection \cite{Tremblay:2014hl}. The problem of detecting multiscale community in networks is cast as the problem of clustering nodes based on graph wavelets features. This allows the introduction of a notion of scale in the analysis of the network and as well as a sort of 'egocentered' view of how a particular node 'sees' the network (see Figure \ref{fig:communities}). Furthermore, the extension of clustering or community detection tasks to large-scale systems generally relies on sampling or randomized strategy where GSP methods can also be very helpful. For example, fast graph-based filtering of random signals can be used to estimate the graph structure, and in particular to approximate eigenvectors that are often crucial in the design of clustering algorithms and other machine learning tasks. One of the initial works in this direction \cite{Boutsidis:2015vp} proposes to use power methods (that can be shown to be related to graph filter operators) to speed up the computation of eigenvectors used in spectral clustering applications. More recently, a fast graph clustering algorithm that is provably as good as spectral clustering has been developed based on random signal filtering techniques \cite{Tremblay:2016vb}. Related ideas have been used in sketching \cite{Chi:2015cy,Gama:2016ui}, data visualization applications on large real-world datasets of millions of nodes \cite{Paratte:2016ua,ChenTFVK:16}, \NEW{or in analysis of dynamic networks \cite{dal2017wavelet}}. These examples provide evidence for the potential benefits  of using  GSP principles in big data applications. 

Finally, the GSP framework can also be used to design architectures to analyze or classify whole graph signals that originally live on irregular structures. In particular, the graph signal processing toolbox has been extensively used to extend convolutional deep learning techniques to data defined on graphs. The convolutional neural network paradigm has been generalized with help of GSP  elements for the extraction of feature descriptors for 3D shapes \cite{Masci:2015ej,Boscaini:2015uea}. A localized spectral network architecture leveraging on localized vertex-frequency analysis has also been proposed in \cite{Bruna:2013vg}, and the use of heat kernels defined in the graph spectral domain has been developed in \cite{Boscaini:2016tx}. While the previous works mostly address the analysis of 3D shapes, convolutional neural networks (CNNs) can actually be extended to many other signals in high-dimensional irregular domains, such as social networks, brain connectomes or words embedding, by reformulation in the context of spectral graph theory. Here, the GSP framework leads to the development of fast localized convolutional filters on graphs \cite{Defferrard:2016vo} along with adapted pooling operators \cite{Khasanova:2017vs}. Unsurprisingly, deep network architectures for graphs signals have been actually tested in various applications domains, such as chemical molecule properties prediction \cite{Duvenaud:2015ww}, classification tasks on social networks \cite{Perozzi:2014ib}, autism spectrum disorder classification \cite{Anirudh:2017tn} or traffic forecasting \cite{Li:2017tn}.


\section{Conclusion}
\label{sec:conclusion}
While recent papers have developed key principles for signal processing of graph signals, and these have shown significant promise for some important applications, there remain significant challenges. On the theoretical front, work to date has focused on results that can be applied to arbitrary graphs. But given the significant differences between the spectral properties of graphs, there is strong current interest in developing tools that can take into consideration the particular characteristics of specific classes of graphs. On the application front, GSP is a good match for datasets exhibiting irregular relationships between samples that can be captured by a graph. However, additional research is needed within each application to further understand the best ways to combine GSP tools with existing techniques in order to achieve significant gains in terms of the metrics of interest for each application. Finally, it is worth mentioning that many of the basic GSP tools described here are available in several Matlab/Python toolboxes: GSPBox \cite{perraudin2014gspbox}, GraSP \cite{Girault-GraSP-17} and PyGSP \cite{pygsp}.

\par\noindent
\NEW{{\bf Acknowledgements:} The authors would like to thank the anonymous reviewers for their constructive comments, which helped improve the final version of the manuscript\NEW{; the authors who kindly helped in the description of their work and permitted reproduction of their figures;} and Dr.~Benjamin Girault (USC) for his careful reading and suggestions, and for his help with some of the figures included in the paper. }

\ifCLASSOPTIONcaptionsoff
  \newpage
\fi



\bibliographystyle{IEEEtran}
\bibliography{./bibtex/bib/IEEEabrv,./bibtex/bib/ProcIEEE-jk,./bibtex/bib/ProcIEEE-pf,./bibtex/bib/ProcIEEE-ao,./bibtex/bib/IEEE-Proc-jm}
%
%
%

%


\begin{IEEEbiographynophoto}{Antonio Ortega}(F'07)  received the Telecommunications Engineering degree from the Universidad Polit\'ecnica de Madrid, Madrid, Spain in 1989 and the Ph.D. in Electrical Engineering from Columbia University, New York, NY in 1994. At Columbia he was supported by a Fulbright scholarship. 

In 1994 he joined the Electrical Engineering department at the University of Southern California (USC), where he is currently a Professor.  He has served as Associate Chair of EE-Systems and director of the Signal and Image Processing Institute at USC.  He is a Fellow of IEEE and EURASIP, and a member of ACM and APSIPA. He has been Chair of the Image and Multidimensional Signal Processing (IMDSP) technical committee, and chair of the SPS  Big Data Special Interest Group.  He has been technical program co-chair of MMSP 1998, ICME 2002, ICIP 2008, PCS 2013, PCS 2018 and DSW 2018, Senior Associate Editor for the IEEE Transactions on Image Processing (IEEE TIP) and the IEEE Signal Processing Magazine, among others.  He is the inaugural Editor-in-Chief of the APSIPA Transactions on Signal and Information Processing, an Associate Editor of IEEE T-SIPN. He currently serves as a member at large of the board of governors of the IEEE SPS. He received the NSF CAREER award, the 1997 IEEE Communications Society Leonard G. Abraham Prize Paper Award, the IEEE Signal Processing Society 1999 Magazine Award, the 2006 EURASIP Journal of Advances in Signal Processing Best Paper Award, the ICIP 2011 best paper award, a best paper award at Globecom 2012, and the 2016 SPM Award. He was a plenary speaker at ICIP 2013, APSIPA ASC 2015 and CAMSAP 2017.  

His research interests are in the areas of signal compression, representation, communication and analysis. His recent work is focusing on distributed compression, multiview coding, error tolerant compression, information representation in wireless sensor networks and graph signal processing. Over 40 PhD students have completed their PhD thesis under his supervision at USC and his work has led to over 400 publications in international conferences and journals, as well as several patents. 

\end{IEEEbiographynophoto}
\begin{IEEEbiographynophoto}{Pascal~Frossard}(S'96-M'01--SM'04--F'18) received the M.S. and Ph.D. degrees, both in electrical engineering, from the Swiss Federal Institute of Technology (EPFL), Lausanne, Switzerland, in 1997 and 2000, respectively. Between 2001 and 2003, he was a member of the research staff at the IBM T. J. Watson Research Center, Yorktown Heights, NY, where he worked on media coding and streaming technologies. Since 2003, he has been a faculty at EPFL, where he heads the Signal Processing Laboratory (LTS4). His research interests include graph signal processing, image representation and coding, visual information analysis, and distributed signal processing and communications.

Dr. Frossard has been the General Chair of IEEE ICME 2002 and Packet Video 2007. He has been the Technical Program Chair of IEEE ICIP 2014 and EUSIPCO 2008, and a member of the organizing or technical program committees of numerous conferences. He has been an Associate Editor of the IEEE TRANSACTIONS ON SIGNAL PROCESSING (2015-), IEEE TRANSACTIONS ON BIG DATA (2015-), IEEE TRANSACTIONS ON IMAGE PROCESSING (2010-2013), the IEEE TRANSACTIONS ON MULTIMEDIA (2004-2012), and the IEEE TRANSACTIONS ON CIRCUITS AND SYSTEMS FOR VIDEO TECHNOLOGY (2006-2011). He is an elected member of the IEEE Multimedia Signal Processing Technical Committee (2004-2007, 2016-), the IEEE Visual Signal Processing and Communications Technical Committee (2006-) and the IEEE Multimedia Systems and Applications Technical Committee (2005-). He has served as Chair of the IEEE Image, Video and Multidimensional Signal Processing Technical Committee (2014-2015), and Steering Committee Chair (2012-2014) and Vice-Chair (2004-2006) of the IEEE Multimedia Communications Technical Committee. He received the Swiss NSF Professorship Award in 2003, the IBM Faculty Award in 2005, the IBM Exploratory Stream Analytics Innovation Award in 2008, the IEEE Transactions on Multimedia Best Paper Award in 2011, and the IEEE Signal Processing Magazine Best Paper Award 2016.
\end{IEEEbiographynophoto}
\begin{IEEEbiographynophoto}{Jelena~Kova\v{c}evi\'c}
(S'88--M'91--SM'96--F'02) received the Dipl. Electr. Eng. degree from
the EE Department, University of Belgrade, Yugoslavia, in 1986, and
the M.S. and Ph.D. degrees from Columbia University, New York, in 1988
and 1991, respectively. From 1991--2002, she was with Bell Labs,
Murray Hill, NJ. She was a co-founder and Technical VP of xWaveforms,
based in New York City and an Adjunct Professor at Columbia
University. In 2003, she joined Carnegie Mellon University, where she
is Hamerschlag University Professor and Head of Electrical and Computer Engineering, Professor of Biomedical Engineering, and was the Director of the Center for Bioimage Informatics at Carnegie Mellon University. Her research interests include wavelets, frames, graphs, and applications to bioimaging and smart infrastructure.

Dr.~Kova\v{c}evi\'{c} coauthored the books Wavelets and Subband Coding
(Prentice Hall, 1995) and Foundations of Signal Processing (Cambridge
University Press, 2014), a top-10 cited paper in the Journal of
Applied and Computational Harmonic Analysis, and the paper for which
A.~Mojsilovi\'{c} received the Young Author Best Paper Award. Her
paper on multidimensional filter banks and wavelets was selected as
one of the Fundamental Papers in Wavelet Theory. She received the
Belgrade October Prize in 1986, the E.I. Jury Award at Columbia
University in 1991, the 2010 CIT Philip L. Dowd Fellowship Award
from the College of Engineering at Carnegie Mellon University and the 2016 IEEE SPS Technical Achievement Award. She is a past Editor-in-Chief of the IEEE Transactions on Image Processing, served as a guest co-editor on a number of special issues and is/was on the editorial boards of several journals.  She was a regular member of the NIH Microscopic Imaging Study Section and served as a Member-at-Large of the IEEE Signal Processing Society Board of Governors.  She is a past Chair of the IEEE Signal Processing Society Bio Imaging and Signal Processing Technical Committee.  She has been involved in organizing numerous conferences.  She was a plenary/keynote speaker at a number of international conferences and meetings.
\end{IEEEbiographynophoto}

\begin{IEEEbiographynophoto}{Jos{\'e} M.~F.~Moura} (S'71--M'75--SM'79--F'94) received the engenheiro electrot\'{e}cnico degree from Instituto Superior T\'ecnico (IST), Lisbon, Portugal, and the M.Sc., E.E., and D.Sc.~degrees in EECS from the Massachusetts Institue of Technology (MIT), Cambridge, MA.  
He is the Philip L.~and Marsha Dowd University Professor at Carnegie Mellon University (CMU). He was on the faculty at IST and has held visiting faculty appointments at MIT and New York University (NYU). He founded and directs a large education and research program between CMU and Portugal, www.cmuportugal.org. 
His research interests are on  data science, graph signal processing, and statistical and algebraic signal and image processing. He has published over 550 papers and holds fourteen patents issued by the US Patent Office. The technology of two of his patents (co-inventor A. Kav\v{c}i\'c) are in over three billion disk drive read channel chips, about 60~\% of all computers sold in the last 13 years worldwide, and was, in 2016, the subject of a 750 million US dollars settlement, the largest university settlement ever in the information technologies area.

Dr. Moura is the 2018 IEEE President Elect, he was the IEEE Technical Activities Vice-President (2016) and member of the IEEE Board of Directors. He served in several other capacities including IEEE Division IX Director, member of several IEEE Boards, President of the IEEE Signal Processing Society(SPS), Editor in Chief for the IEEE Transactions in Signal Processing, interim Editor in Chief for the IEEE Signal Processing Letters. 
Dr. Moura has received several awards, including  the Technical Achievement Award and the Society Award from the IEEE Signal Processing. In 2016, he received the CMU College of Engineering Distinguished Professor of Engineering Award. He is a Fellow of the IEEE, a Fellow of the American Association for the Advancement of Science (AAAS), a corresponding member of the Academy of Sciences of Portugal, Fellow of the US National Academy of Inventors, and a member of the US National Academy of Engineering.
\end{IEEEbiographynophoto}


\begin{IEEEbiographynophoto}{Pierre Vandergheynst} received the M.S. degree in physics and the Ph.D. degree in mathematical
physics from the Universit\'{e} catholique de Louvain, Louvain-la-Neuve, Belgium, in 1995 and 1998, respectively. From 1998 to 2001, he was a Postdoctoral Researcher with the Signal Processing Laboratory, Swiss Federal Institute of Technology (EPFL), Lausanne, Switzerland. He was Assistant Professor at EPFL (2002-2007), where he is now a Full Professor of Electrical Engineering and, by courtesy, of Computer and Communication Sciences. As of 201, Prof. Vandergheynst serves as EPFL’s Vice-President for Education.

His research focuses on harmonic analysis, sparse approximations and mathematical data processing in general with applications covering signal, image and high dimensional data processing, computer vision, machine learning, data science and graph-based data processing.

He was co-Editor-in-Chief of Signal Processing (2002-2006), Associate Editor of the IEEE Transactions on Signal Processing (2007-2011), the flagship journal of the signal processing community, and currently serves as Associate Editor of Computer Vision and Image Understanding and SIAM Imaging Sciences. He has been on the Technical Committee of various conferences, serves on the steering committee of the SPARS workshop and was co-General Chairman of the EUSIPCO 2008 conference. 

Pierre Vandergheynst is the author or co-author of more than 70 journal papers, one monograph and several book chapters. He has received two IEEE best paper awards. Professor Vandergheynst is a laureate of the Apple 2007 ARTS award and of the 2009-2010 De Boelpaepe prize of the Royal Academy of Sciences of Belgium.\end{IEEEbiographynophoto}





\end{document}